\documentclass[journal]{IEEEtran}
\usepackage{amsmath,amsfonts}
\usepackage{algorithmic}
\usepackage{algorithm}
\usepackage{amsmath}
\usepackage{array}
\usepackage{subfigure}
\usepackage{textcomp}
\usepackage{stfloats}
\usepackage{url}
\usepackage{cleveref}
\usepackage{verbatim}
\usepackage{graphicx}
\usepackage{multirow}
\usepackage{titlesec}
\usepackage{makecell}
\usepackage{epstopdf}
\usepackage{cite}
\usepackage{xcolor}

\begin{document}
\title{SAR-GS: Gaussian Splatting based SAR Images Rendering and Target Reconstruction}

\author{Aobo Li,~\IEEEmembership{Graduate Student Member,~IEEE,} Zhengxin Lei,~\IEEEmembership{Graduate Student Member,~IEEE,} \\ Jiangtao Wei,~\IEEEmembership{Member,~IEEE,} Feng Xu,~\IEEEmembership{Senior Member,~IEEE}
\thanks{The authors are with the Key Laboratory for Information Science of Electromagnetic Waves (MoE), Fudan University, Shanghai 200433, China (Corresponding author: Jiangtao Wei, e-mail: jtwei21@m.fudan.edu.cn).}}


\maketitle

\begin{abstract}
    Three-dimensional target reconstruction from synthetic aperture radar (SAR) imagery is crucial for interpreting complex scattering information in SAR data. However, the intricate electromagnetic scattering mechanisms inherent to SAR imaging pose significant reconstruction challenges. Inspired by the remarkable success of 3D Gaussian Splatting (3D-GS) in optical domain reconstruction, this paper presents a novel  SAR Differentiable Gaussian Splatting Rasterizer (SDGR) specifically designed for SAR target reconstruction.
    Our approach combines Gaussian splatting with the Mapping and Projection Algorithm to compute scattering intensities of Gaussian primitives and generate simulated SAR images through SDGR. Subsequently, the loss function between the rendered image and the ground truth image is computed to optimize the Gaussian primitive parameters representing the scene, while a custom CUDA gradient flow is employed to replace automatic differentiation for accelerated gradient computation. Through experiments involving the rendering of simplified architectural targets and SAR images of multiple vehicle targets, we validate the imaging rationality of SDGR on simulated SAR imagery. Furthermore, the effectiveness of our method for target reconstruction is demonstrated on both simulated and real-world datasets containing multiple vehicle targets, with quantitative evaluations conducted to assess its reconstruction performance. Experimental results indicate that our approach can effectively reconstruct the geometric structures and scattering properties of targets, thereby providing a novel solution for 3D reconstruction in the field of SAR imaging.
\end{abstract}

\begin{IEEEkeywords}
Synthetic aperture radar(SAR), Gaussian splatting,differentiable SAR rasterizer, mapping and projection algorithm, SAR target reconstruction.
\end{IEEEkeywords}

\section{Introduction}
\IEEEPARstart{S}{ynthetic} Aperture Radar (SAR), as an active microwave imaging system, has emerged as a pivotal technological tool in Earth remote sensing due to its all-weather and day-night imaging capabilities, demonstrating unique application value in military reconnaissance and environmental monitoring\cite{APreliminaryStudyonSARAdvancedInformationRetrievalandSceneReconstruction,1}. Unlike optical camera imaging principles,  the active sensing mechanism of SAR acquires target information through electromagnetic wave scattering interactions with terrain features \cite{1717717,4909462}. This results in pronounced geometric sensitivity in SAR imagery, where identical targets exhibit substantial appearance variations under different viewing angles, frequency bands, and polarization configurations \cite{2014Microwave}. While this distinctive characteristic provides unique advantages in multidimensional information acquisition, it poses formidable challenges for SAR image interpretation and three-dimensional (3D) geometric reconstruction tasks. Traditional 3D reconstruction techniques critically depend on profound understanding of complex scattering mechanisms and imaging geometry.


Traditional SAR three-dimensional reconstruction methodologies predominantly rely on interferometric and tomographic principles. Established techniques like Interferometric SAR (InSAR)\cite{42,43} and Tomographic SAR (TomoSAR)\cite{46} have demonstrated practical success in specific scenarios. For instance, InSAR has been extensively employed for surface elevation mapping through phase difference analysis\cite{7},  while Zhu et al. \cite{48} achieved three-dimensional reconstruction of Berlin, Germany, through TomoSAR processing of over 450 TerraSAR-X images with 1-meter resolution. Knaell's seminal 3D SAR concept\cite{45} and subsequent Array InSAR developments\cite{47} improved elevation resolution through synthetic aperture synthesis, they remain constrained by geometric inversion paradigms rather than leveraging structural/textural semantics. Recent advancements attempt to overcome these limitations through multimodal fusion. Notably, Ding et al.\cite{Ding2019SyntheticAR} proposed a paradigm-shifting framework integrating TomoSAR with scattering characteristics and visual semantic features extracted from 2D SAR images. This approach employs learned semantic constraints as regularization terms, effectively reducing required acquisition orbits while enhancing reconstruction fidelity in complex urban environments. Meanwhile, in the field of microwave vision \cite{R23225}, Xu et al. \cite{yuyi} proposed a semantic electromagnetic scattering modeling method for radar sensing. This method provides a novel solution and technical pathway for solving electromagnetic inverse problems.

Traditional 3D reconstruction techniques critically rely on an in-depth understanding of complex scattering mechanisms and imaging geometry. In contrast, differentiable scene modeling approaches offer novel solutions for decoding intricate scattering information embedded in SAR imagery, enabling the simultaneous and accurate recovery of both geometric structures and scattering properties of observed scenes. Wei et al. \cite{Wei} achieved high-fidelity SAR simulation by integrating differentiable ray tracing (DRT) with physical scattering models (KA+SPM), successfully fitting the scattering material parameters of object surfaces through inverse gradient propagation. Li et al. \cite{Li} proposed a gradient descent-based method to learn the parameters of scattering models, reliably estimating scene parameter maps while preserving texture features. The simulated multi-angle SAR data derived from these maps exhibited strong radiometric consistency with real measurements.

Concurrently, driven by the proliferation of large-scale datasets, data-driven neural network architectures tailored for SAR applications have emerged extensively. Qin et al. \cite{10579764} reformulated the reconstruction task as a sequence-to-sequence learning problem by embedding electromagnetic scattering priors into a Transformer model. Fu et al. \cite{9926979} proposed a differentiable SAR renderer, enabling joint inversion of 3D geometric structures and facet scattering intensities. Lei et al. \cite{SARNERF}extended Neural Radiance Fields (NeRF) \cite{nerf} to the SAR domain via SAR-NeRF, achieving novel view synthesis while preserving radar imaging physics. Other notable innovations include cross-modal transformation techniques (e.g., SAR-optical generative adversarial networks leveraging pre-trained optical networks \cite{85}), resolution-enhancing CNNs for limited observation orbits \cite{86}, and physically constrained complex-valued CNN methods for building façade reconstruction\cite{87}.

Although the imaging mechanisms of optical systems differ from those of SAR, data-driven 3D reconstruction methods in optical imaging can provide valuable insights. Current data-driven 3D reconstruction algorithms in optical imaging primarily adopt two supervision paradigms: 3D supervision relying on explicit 3D ground-truth models and 2D supervision. Among these, image-based reconstruction algorithms have garnered increasing attention from researchers due to their reduced dependence on high-quality 3D data. However, image-based 3D reconstruction faces a fundamental challenge—the non-differentiable nature of traditional rendering pipelines, where discrete rasterization obstructs gradient backpropagation. To address this, researchers have developed various gradient approximation techniques, including differentiable filtering (OpenDR \cite{97}), soft rasterization (SoftRas \cite{softras}), differentiable ray tracing (DIRT \cite{DIRT}), differentiable interpolation-based rendering (DIBR \cite{DIBR}), reparameterized rendering (Redner \cite{REDNER}), non-local gradient propagation (Neural 3D Mesh Renderer \cite{98}), and pixel-level barycentric coordinate derivatives \cite{99}. These techniques enable continuous optimization of geometric parameters.

The integration of explicit 3D representations with differentiable rendering techniques has recently led to the emergence of 3D Gaussian Splatting (3D-GS)\cite{kerbl3Dgaussians}, a method centered on 3D Gaussian-distributed point cloud modeling. The 3D-GS pipeline consists of two key stages: scene modeling and differentiable rendering.
During the scene modeling phase, the system adaptively optimizes the parameters of millions of anisotropic Gaussian primitives—including their spatial positions, covariance matrices, and color attributes—using multi-view image sequences and sensor pose data. This optimization leverages point cloud density control and gradient descent algorithms to accurately reconstruct both geometric structures and appearance characteristics.
In the rendering phase, 3D-GS employs a Gaussian splatting strategy, where pixel colors are synthesized via alpha-blending. Its key advantages lie in its breakthrough rendering efficiency and novel explicit scene representation, while maintaining high-fidelity scene reconstruction accuracy.


Inspired by 3D-GS, this paper proposes a SAR image-based target reconstruction algorithm (SAR Gaussian Splatting, SAR-GS). The novel SAR Differentiable Gaussian Rasterizer (SDGR) inherently integrates the forward SAR imaging process with the inverse target reconstruction algorithm. By combining the Mapping and Projection Algorithm \cite{4241254} for SAR imaging with Gaussian splatting techniques, we establish a unified framework that bridges SAR imaging and target reconstruction. The main contributions of this work are as follows:
\begin{enumerate}
    \item Novel SAR differentiable Gaussian rasterizer: By integrating MPA with the gaussian splattering methods, we introduce a novel SAR differentiable Gaussian rasterizer.
    \item 3D target reconstruction: We design the inverse gradients of SDGR by backpropagating from the rendered images to the electromagnetic scattering properties and geometric configurations of the target Gaussians. This is achieved by deriving the first-order gradients of the differentiable Gaussian rasterization imaging algorithm, thereby formulating the inverse gradients for SDGR.
    \item Extensive validation and evaluation: We comprehensively validated the effectiveness of SDGR through both forward simulation experiments and target reconstruction tasks using both simulated and real-measured datasets. Furthermore, ablation studies were conducted to investigate the impacts of initial point clouds and densification operations on reconstruction performance.
\end{enumerate}

The structure of this paper is organized as follows: Section II presents the overall framework and technical details of the SAR-GS method, including the forward rendering incorporating Mapping and Projection Algorithm and Gaussian Splatting approaches, along with backward gradient derivation. Section III provides experimental results and comparative analysis of target reconstruction based on both simulated datasets and real-measured datasets. Section IV concludes the paper with a comprehensive summary and discusses potential future research directions in this field.

\section{SAR-GS}
\begin{figure*}[htbp]
    \centering
    \includegraphics[width=1\linewidth]{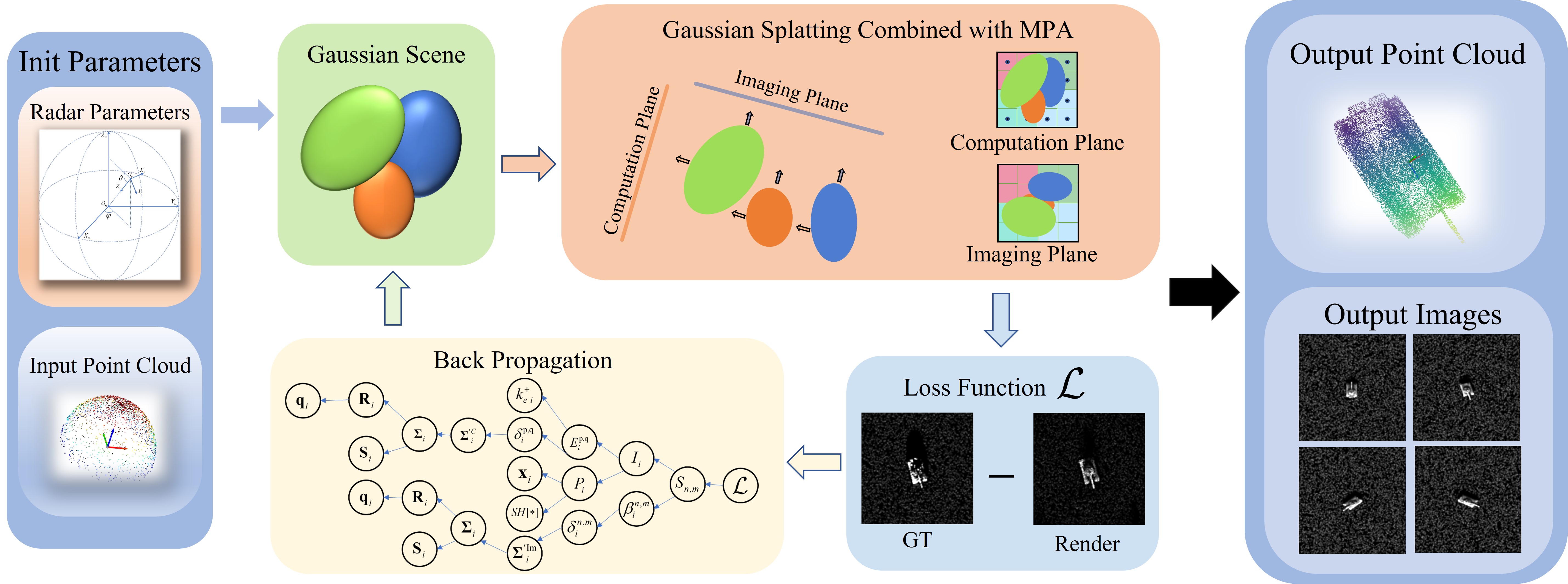}
    \caption{The general workflow of a target reconstruction algorithm based on SAR Gaussian Splatting.}
    \label{SARGSflow}
\end{figure*}

This section presents the overall algorithmic workflow of SAR-GS, with the corresponding flowchart illustrated in Fig.\ref{SARGSflow}. The algorithm consists of two key processes: forward rendering and backward gradient propagation.

In the forward rendering process, Gaussian primitives within the scene are projected onto the computational plane via Gaussian splatting operations. The computational plane is discretized into grids, and rays are emitted from grid centers to simulate the radar beam sampling process along the azimuth direction, while also modeling the propagation and scattering intensity attenuation of radar waves among Gaussian primitives. The scattering intensity of each Gaussian primitive is determined by radar parameters, its spatial coordinates, and scattering coefficients. Subsequently, these Gaussian primitives are reprojected onto the imaging plane through Gaussian splatting—where the scattering intensities of primitives aligned in the same range and azimuth directions are synthesized to form the backscattered energy of range-azimuth pixels, ultimately generating a complete SAR image.

In the backward gradient propagation stage, the gradients of pixel energy with respect to Gaussian primitive parameters are computed, and a gradient descent-based backpropagation algorithm is employed to optimize the geometric attributes and scattering coefficients of the Gaussian primitives. This enables the task of refining Gaussian primitive parameters to accurately represent the scattering scene based on SAR image pixels. 

\subsection{Gaussian Splatting Combined with MPA}
The framework primarily comprises three primary components:
\begin{itemize}
    \item Mapping and Projection Algorithm: Based on radar parameters and the geometric parameters of Gaussian primitives (position $x$ and covariance matrix $\Sigma$), Gaussian primitives are projected onto predefined computation plane and imaging plane. This yields their corresponding 2D coordinates ($x_c,x_i$) and 2D covariance matrices ($\Sigma^{\prime}_c$,$\Sigma^{\prime}_i$ ) on these planes.
    \item Scattering Intensities Computation: The scattering intensities $I_i$ of Gaussian primitives are calculated sequentially along depth dimensions using depth-indexed Gaussian lists generated by the SA-GS framework.
    \item Image Formation: Scattering intensities $I_i$ of Gaussian primitives aligned within the same range direction are cumulatively integrated to synthesize the backscattered energy for individual range-azimuth pixels, thereby generating the final SAR image.
\end{itemize}

\subsubsection{Mapping and Projection Algorithm}
\begin{figure}[h]
    \centering
    \includegraphics[width=1\linewidth]{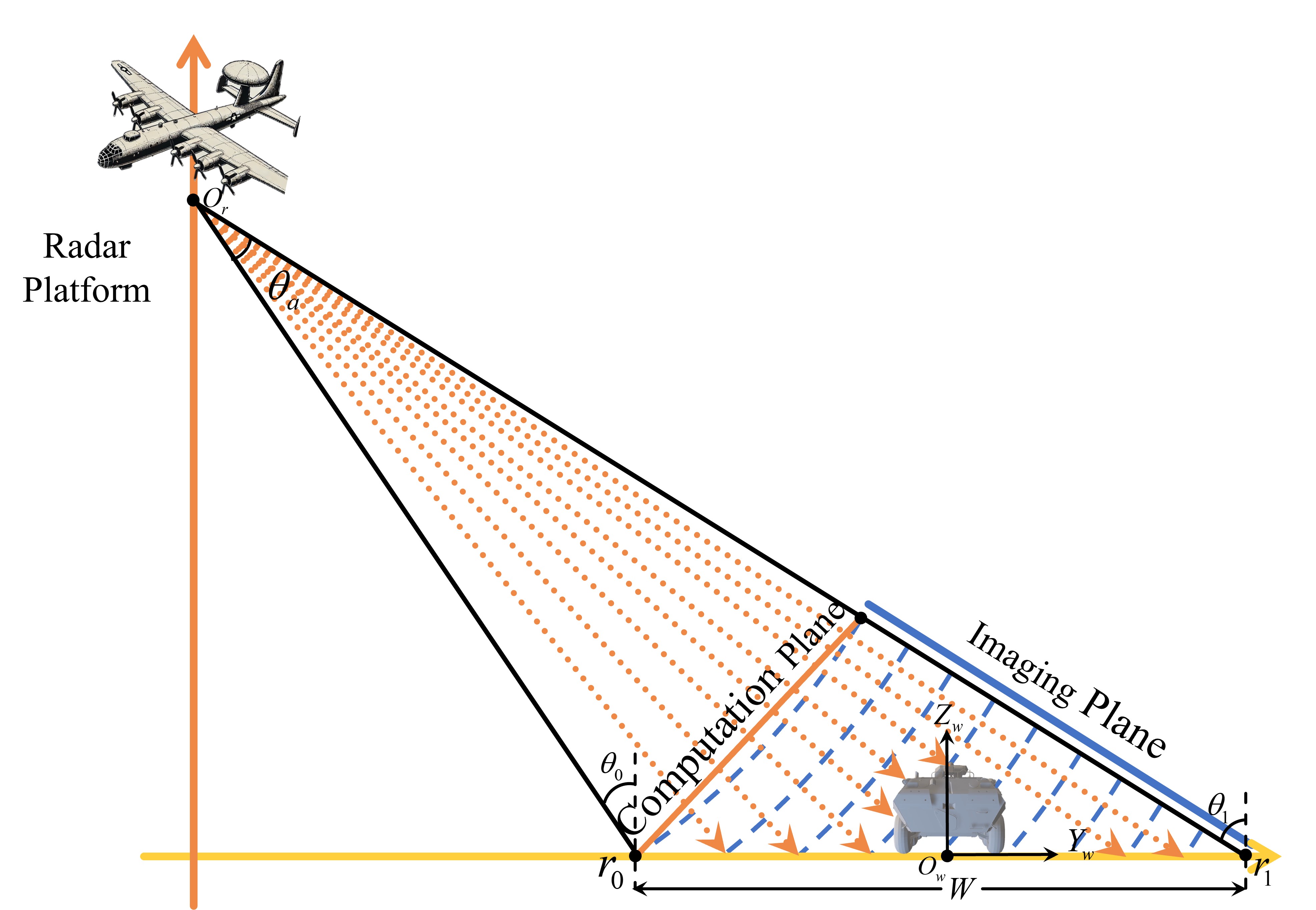}
    \caption{Mapping Projection Algorithm (Single Azimuth Direction)}
    \label{MPA2.png}
\end{figure}

SAR achieves two-dimensional high-resolution imaging through pulse compression and synthetic aperture techniques, as shown in Fig. \ref{MPA2.png}. The platform flies at an altitude $h$ along the $X_\text{r}$-axis, continuously transmitting signals towards the ground and receiving echo reflections from ground targets. Throughout the process, the radar antenna maintains a fixed viewpoint (typically a side view). Let the origin $O_\text{w}$ of the world coordinate system denote the center of the illuminated area. $\theta_\text{r}$ and $\theta_\text{a}$ are the vertical and azimuth beamwidths, respectively, $W$ is the swath width of the imaging area.
\begin{figure}[h]
    \centering
    \includegraphics[width=1\linewidth]{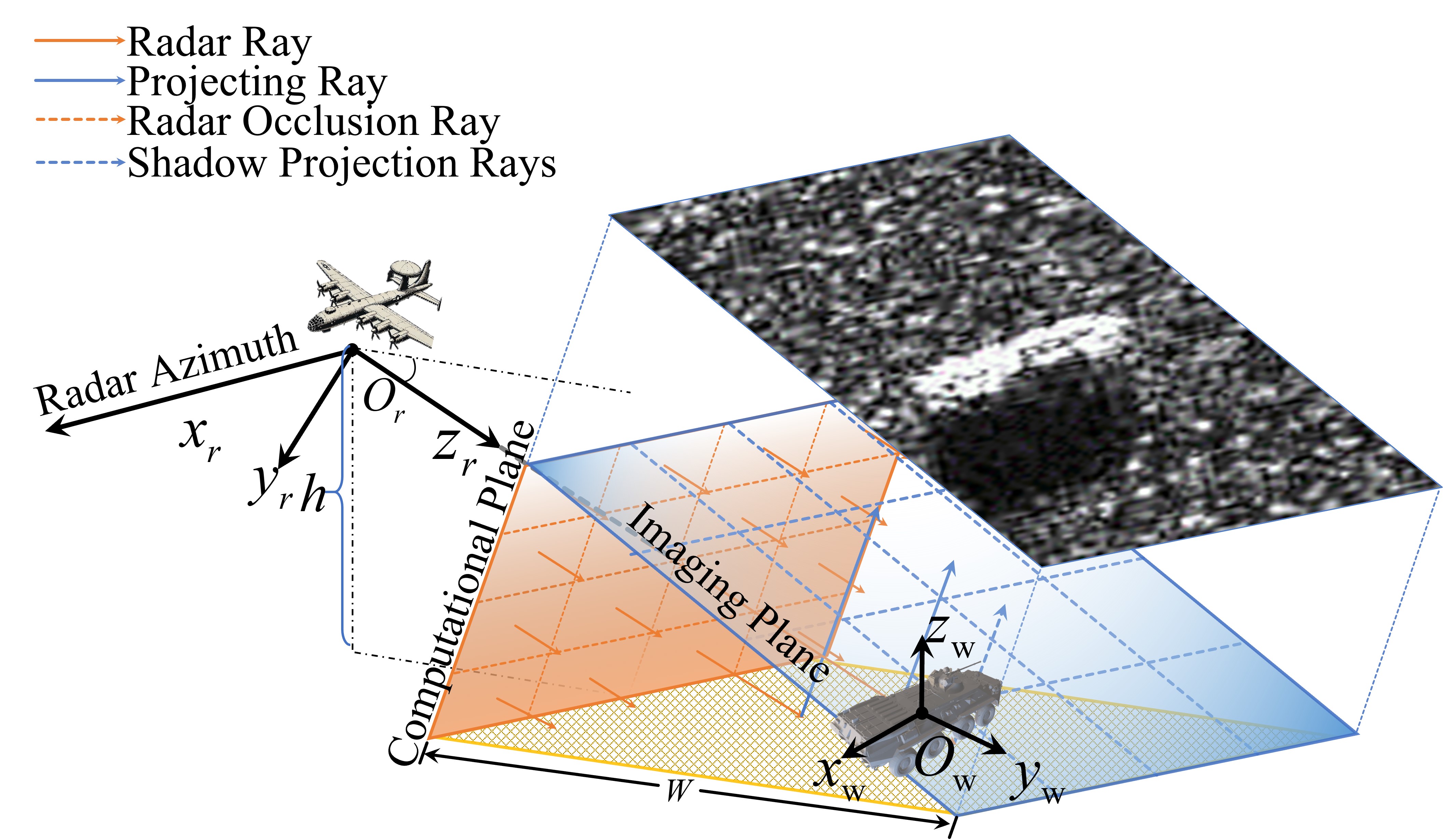}
    \caption{Mapping and Projection Algorithm.The airborne radar travels along the $O_\text{r}X_\text{r}$ direction (azimuth direction) and emits radar beams along the $O_\text{r}Z_\text{r}$  direction (range direction).}
    \label{MPA.png}
\end{figure}

\begin{figure}
    \centering
    \includegraphics[width=0.5 \linewidth]{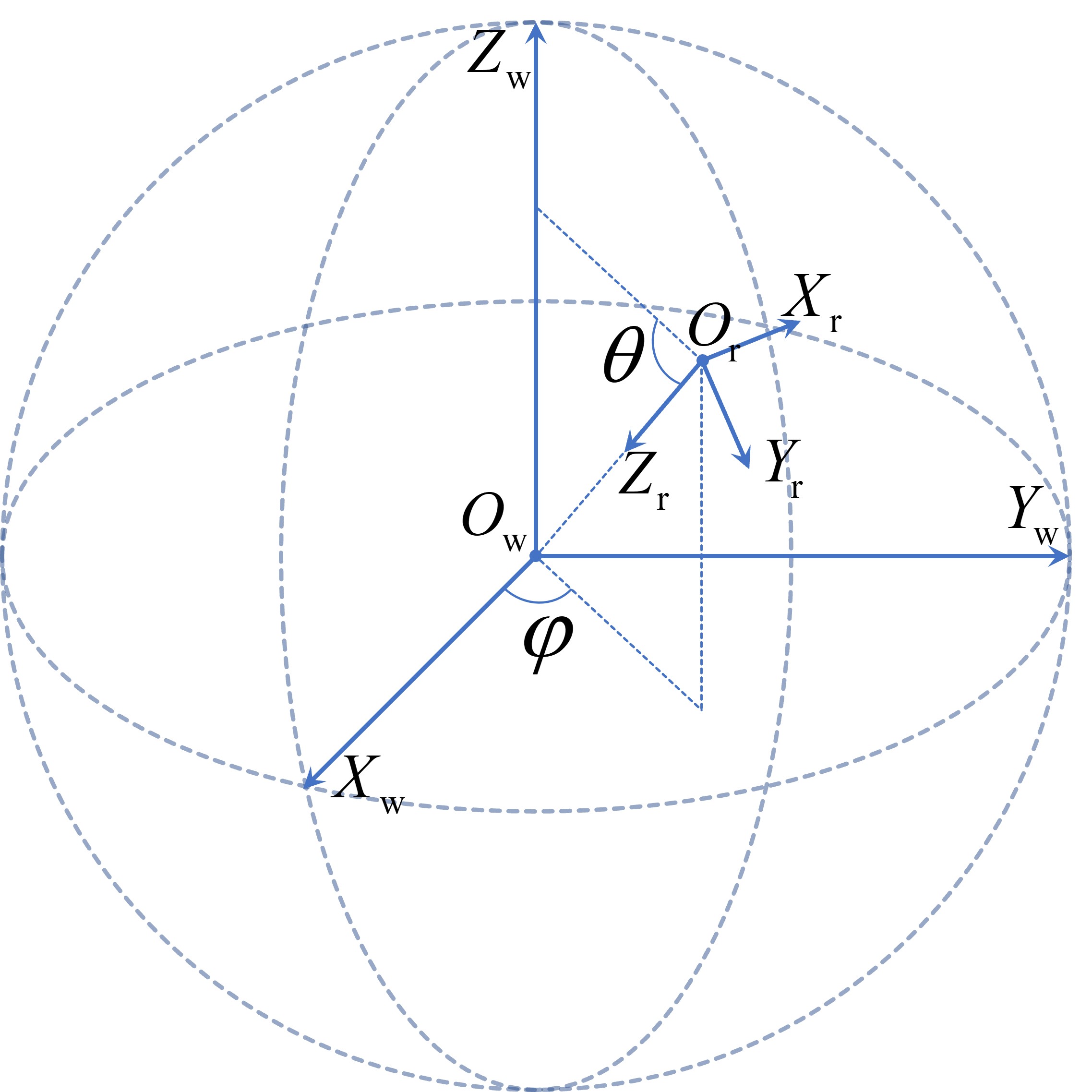}
    \caption{Schematic Diagram of the Transformation between Radar Coordinate System and World Coordinate System, where $O_\text{w}-X_\text{w}Y_\text{w}Z_\text{w}$ represents the World Coordinate System and $O_\text{r}-X_\text{r}Y_\text{r}Z_\text{r}$ denotes the Radar Coordinate System. $\phi$ is the radar azimuth angle, and $\theta$ is the radar elevation angle.}
    \label{fig：transform}
\end{figure}

A single cross-sectional plane along the azimuthal direction is defined as the incidence plane, establishing a polar coordinate system $(r,\theta)$ with the radar position as the origin. Here, $\theta$ denotes the radar incidence angle, and $r$ represents the slant range. By determining the sampling range of radar-received pulse echoes and the variation range of the incidence angle, the radar imaging domain is defined as $ r \in [r_0,r_1]$ and $\theta \in [\theta_0 , \theta_1]$ Consider a grid cell within the imaging space, corresponding to a voxel $\text{d}v$ with dimensions $\text{d}x$, $\text{d}r$, and $r\text{d}\theta$. According to radiative transfer theory, the backscattering intensity $I_s$  per unit area, when an incident wave ropagates through a single voxel, can be expressed as :
\begin{subequations}
\begin{align}           
I_{s}\left(x,r,\theta\right)=E^{+}\left(x,r,\theta\right)P\left(x,r,\theta\right)E^{-}\left(x,r,\theta\right)I_{i}\mathrm{d}r \label{I_MPA}\\
E^{+}\left(x,r,\theta\right)=\exp\left[-\int_{r_{0}}^{r}\mathrm{d}r^{\prime}k_{e}^{+}\left(x,r^{\prime},\theta\right)\right] \label{E+MPA}\\
E^{-}\left(x,r,\theta\right)=\exp\left[-\int_{r}^{r_{0}}\mathrm{d}r^{\prime}k_{e}^{-}\left(x,r^{\prime},\theta\right)\right] \label{E-MPA}
\end{align}
\end{subequations}
The terms $E^+$ and $E^-$ denote the cumulative attenuation coefficients in the forward and backward directions, respectively. The function PP represents the scattering coefficient of the voxel, while $k^+_e\left(x,r,\theta\right)$ and $k^-_e \left(x,r,\theta\right)$ correspond to the extinction coefficients in the forward and backward directions. The contribution of scattered energy is derived from the product of the scattering intensity of the scatterer and its effective penetration area, expressed as:
\begin{equation}
S\left(x,r\right)\mathrm{d}x=\int_{\theta_0}^{\theta_1}I_s\left(x,r,\theta\right)r\mathrm{d}x\mathrm{d}\theta \label{S}
\end{equation}
By substituting the above equations into \Cref{I_MPA,E+MPA,E-MPA,S}, the scattered energy of a single pixel in the SAR image can be formulated as follows:
\begin{equation}\begin{aligned}S_{i,j}&=\int\int S\left(x,r\right)\mathrm{d}x\mathrm{d}r\\&=\int_{x_{i}}^{x_{i+1}}\int_{r_{j}}^{r_{j+1}}\int_{\theta_{0}}^{\theta_{1}}\exp\left[-\int_{r_{0}}^{r}dr^{\prime}k_{e}^{+}(x,r^{\prime},\theta)\right]\cdot\\&P\left(x,r,\theta\right)\exp\left[-\int_{r}^{r_{0}}dr^{\prime}k_{e}^{-}\left(x,r^{\prime},\theta\right)\right]r\mathrm{d}\theta\mathrm{d}r\mathrm{d}x\end{aligned}\end{equation}
The SAR-GS algorithm involves projecting the Gaussian ellipsoids constituting the scene onto both computation plane and imaging plane to complete the splatting operation, with the projection process being a coordinate transformation procedure centered around transformation matrices.
First, we assume that the target is located at the origin of the world coordinate system, as illustrated in the Fig. \ref{fig：transform}. The transformation matrix $\textbf{W}$ from the world coordinate system to the radar coordinate system can be expressed as:
\begin{equation}
\label{transformation matrix}
\mathbf{W}=\left[ \begin{matrix}
   {{\mathbf{R}}_{r}} & {{\mathbf{T}}_{r}}  \\
   \mathbf{0} & 1  \\
\end{matrix} \right],
\end{equation}
\begin{equation}
    \mathbf{R}_{r} = \begin{bmatrix}
        - \sin\varphi & -\cos\varphi & 0 \\
        -\sin\theta \cos\varphi & \sin\theta\sin\varphi & \cos\theta \\
        -\cos\theta\cos\varphi & \cos\theta\sin\varphi & -\sin\theta \\
        \end{bmatrix},
\end{equation}

where $\textbf{R}_\text{r}$ represents the rotation matrix from the World Coordinate System to the Radar Coordinate System and $\mathbf{T}_{r}$ represents the radar's offset vector.

A 3D Gaussian is a Gaussian probability density distribution in three-dimensional space, which describes the probability density of a point relative to a certain central point. The formula is as follows:
\begin{equation}
    G(\mathbf{x}) = \frac{1}{\sqrt{(2\pi)^3 |\mathbf{\Sigma}|}} e^{-\frac{1}{2} (\mathbf{x} - \mathbf{\mu})^T \mathbf{\Sigma}^{-1} (\mathbf{x} - \mathbf{\mu})}
\end{equation}
where the covariance matrix $\mathbf{\Sigma}$ characterizes the extent of diffusion of the Gaussian , and the mean vector $\mathbf{\mu}$ denotes the position of the Gaussian's center in space.

For mapping coordinates from the radar coordinate system to a two-dimensional plane, we approximate the orthogonal projection matrix using the Jacobian matrix$\mathbf{J}$. The covariance matrix of the 3D Gaussian projected in the image plane can be represented as: 
\begin{equation}
    \Sigma^\prime = \mathbf{J}^{\top} \mathbf{R}^{\top}_\text{r} \mathbf{\Sigma} \mathbf{R}_\text{r} \mathbf{J}.
    \label{Sigmaprime}
\end{equation}
The coordinate transformation of point clouds from the world coordinate system to the radar coordinate system is represented as:
\begin{equation}
    \mathbf{x}_\text{r} = \mathbf{W} \mathbf{x}_\text{w} + \mathbf{T}_\text{r}.
    \label{xprime}
\end{equation}
The corresponding 2D coordinates $\mathbf{x}_C,\mathbf{x}_I$ on both the computation plane and imaging plane can be obtained, with the third-dimensional value serving as depth for sorting purposes.The projection matrices for the computation plane $\mathbf{P}_{C}$ and imaging plane $\mathbf{P}_{I}$ are defined as follows:
\begin{equation}
    \mathbf{P}_{C} = \begin{bmatrix}
        \frac{2}{\Delta A *N_{A}} & 0 & 0 & 0\\
        0 & \frac{2}{{\Delta R} * N_{R} * \tan\theta} & 0 & 0 \\
        0 & 0 & 0 & 0 \\
        0 & 0 & 1 & 0 \\
        \end{bmatrix},
    \label{x_C}
\end{equation}
\begin{equation}
    \mathbf{P}_{I} = \begin{bmatrix}
        2 /(\Delta A *N_{A}) & 0 & 0 & 0\\
        0 & 0 & \frac{2}{{\Delta R} * N_{R} * \tan\theta} &-\frac{2h}{\Delta R*N_{R}\sin\theta}\\
        0 & 0 & 0 & 0 \\
        0 & 1 & 0 & 0 \\
        \end{bmatrix}.
    \label{x_i}
\end{equation}
where $\theta$ is the grazing angle, $\Delta A$ and $\Delta_R$ represent the azimuth and range resolution respectively, and $N_{A}$ and $N_{R}$ denote the number of pixels in azimuth and range directions of the image.

\subsubsection{Scattering Intensities Computation}
Inspired by the 3D-GS optical camera model, this study derives the scattering intensity of each Gaussian primitive by modeling radar beam attenuation during propagation between Gaussian primitives through integration with the MPA (Multi-Phase Approximation) method. The computational projection plane is uniformly partitioned into a grid of $N\times M$ cells, with radar beams simulated by emitting rays from the center of each grid cell to penetrate Gaussian primitives. After obtaining the 2D covariance matrices and planar coordinates for both computational and imaging projection planes, the scattering intensity of Gaussian primitives is calculated via a weighted summation of 2D covariances, the distance from the ray center to planar coordinates, and the backscattering coefficient.

Consider the backscattered intensity $I_i$ of the $i$-th Gaussian primitive residing at the $(p,q)$-th ray in the imaging space. According to radiative transfer theory, the backscattered intensity $I_i$ , when the incident wave $I_0$ propagates through this Gaussian primitive, can be formulated as:
\begin{equation}
\begin{aligned}
I_i = &\sum_{p=1}^N \sum_{q=1}^M \left( \prod_{j=1}^{i-1} P_j \cdot E^{+p,q}_j E^{-p,q}_j \right) 
\left( 1 - E^{+p,q}_i P_i E^{-p,q}_i \right) \\
= &\sum_{p=1}^N \sum_{q=1}^M \left( 1 - \exp \left( -(k^+_{e_i} + k^-_{e_i}) \delta_i^{p,q} \right) P_i \right) 
\prod_{j=1}^{i-1} P_j \\
&\cdot \exp \left( -\sum_{j=0}^{i-1} (k^+_{e_j} + k^-_{e_j}) \delta_j^{p,q} \right) P_j,
\end{aligned}
\end{equation}
where, $\delta^{p,q}_{j}$ is defined as follows:
\begin{equation}
    \delta^{p,q}_{j}={{\left[ \begin{matrix}
   \Delta x_{j}^{p,q}  \\
   \Delta y_{j}^{p,q}  \\
\end{matrix} \right]}^{T}}\Sigma _{}^{\prime }\left[ \begin{matrix}
   \Delta x_{j}^{p,q}  \\
   \Delta y_{j}^{p,q}  \\
\end{matrix} \right],
\end{equation}
where, $E^+$ and $E^-$ represent the cumulative total attenuation of the beam along the forward and backward scattering paths, respectively. The function $P$ determines the scattering characteristics of this Gaussian . The parameters $k{_{e}^{+}}$ and $k{_{e}^{-}}$ represent the forward and backward scattering coefficients, respectively. Since a Gaussian can be penetrated by multiple electromagnetic beams, it is necessary to compute a weighted sum of multiple beams.

\subsubsection{Imaging}
By projecting the Gaussian primitives onto the imaging plane, the scattering energy $S_{n,m}$ for the range pixel, $(n,m)$ is determined by the scattering intensities of the Gaussians that cover that range pixel.
\begin{equation}
  \begin{aligned}
      {{S}_{n,m}}&=\sum\limits_{i\in {{\mathcal{G}}^{n,m}}}{\beta _{i}^{n,m}}{{I}_{i}}\\
      &=\sum\limits_{i\in {{\mathcal{G}}^{n,m}}}{\exp (-{{\left[ \begin{matrix}
   \Delta x_{j}^{n,m}  \\
   \Delta y_{j}^{n,m}  \\
\end{matrix} \right]}^{T}}\Sigma _{Ima}^{\prime }\left[ \begin{matrix}
   \Delta x_{j}^{n,m}  \\
   \Delta y_{j}^{n,m}  \\
\end{matrix} \right]){{I}_{i}}},
  \end{aligned}
\end{equation}
where $I_i$ is the scattering intensity of the $i$-th Gaussian primitive contributing to the pixel $(n,m)$.

\subsection{Backward Gradient Propagation}
\begin{figure}[h]
    \centering
    \includegraphics[width=1\linewidth]{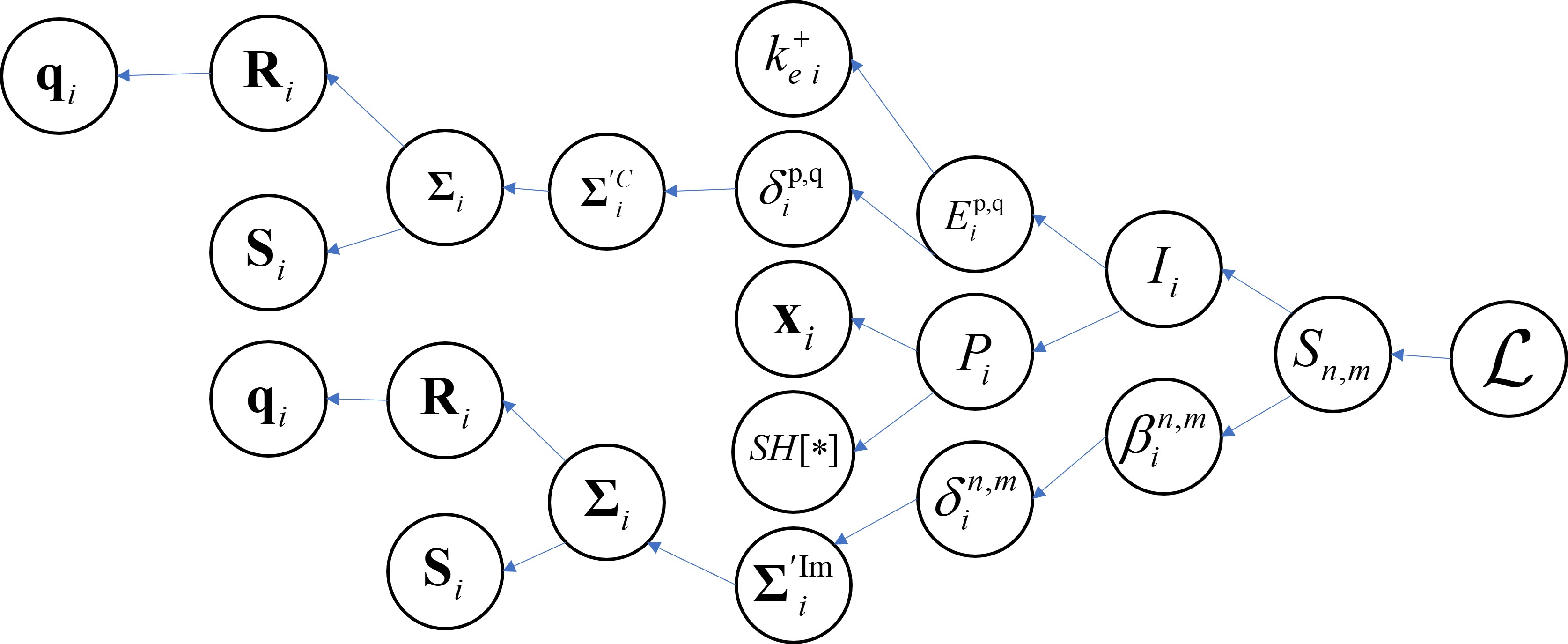}
    \caption{Gradient Flow Diagram}
    \label{grad}
\end{figure}

Building upon the acceleration method employed in 3D-GS\cite{kerbl3Dgaussians}, which defines derivatives within CUDA functions, we similarly accelerated the backward gradients of SDGR by replacing PyTorch’s automatic differentiation with CUDA-based implementations. This approach necessitates deriving the reverse gradients for each component of SDGR, and the gradient backpropagation is illustrated in the accompanying Fig. \ref{grad}.
\subsubsection{Gradient of Imaging Processs}
During the imaging process, two primary components are involved: the gradient of the SAR image pixels with respect to the scattering intensity, and the gradient of the SAR image pixels with respect to the Gaussian kernel contributions. The gradient of the loss function $\mathcal{L}$ with respect to the scattering intensity is given by 
\begin{equation}
\begin{aligned}
    \frac{\text{d}\mathcal{L}}{\text{d}{{I}_{i}}}&=\sum\limits_{n=1}^{H}{\sum\limits_{m=1}^{W}{\frac{\text{d}\mathcal{L}}{\text{d}{{S}_{n,m}}}\frac{\text{d}{{S}_{n,m}}}{\text{d}{{I}_{i}}}}}\\
    &=\underset{n=1}{\overset{H}{\mathop{\sum }}}\,\underset{m=1}{\overset{W}{\mathop{\sum }}}\,\frac{\text{d}\mathcal{L}}{\text{d}{{S}_{n,m}}}\exp (-\delta^{n,m}_{i}),
\end{aligned}
\end{equation}where, $\delta^{n,m}_{j}$ is defined as follows:
\begin{equation}
    \delta^{n,m}_{i}={{\left[ \begin{matrix}
   \Delta x_{i}^{n,m}  \\
   \Delta y_{i}^{n,m}  \\
\end{matrix} \right]}^{T}}\Sigma _{Ima}^{\prime }\left[ \begin{matrix}
   \Delta x_{i}^{n,m}  \\
   \Delta y_{i}^{n,m}  \\
\end{matrix} \right].
\end{equation}
And the gradient of the loss function $\mathcal{L}$ with respect to the Gaussian contribution , is given by 
\begin{equation}
    \frac{\text{d}\mathcal{L}}{\text{d}\beta _{i}^{n,m}}=\frac{\text{d}\mathcal{L}}{\text{d}{{S}_{n,m}}}{{I}_{i}}.
\end{equation} were, $\frac{\text{d}\mathcal{L}}{\text{d}{{S}_{n,m}}}$ represents the gradient of the loss function with respect to the image pixel $(n,m)$, returned by PyTorch's automatic differentiation.

Assuming the two-dimensional covariance matrix as ${{\Sigma }^{\prime }}=\left[ \begin{matrix}
   a & b  \\
   b & c  \\
\end{matrix}\right]$
, the partial derivatives of the loss function with respect to each of its components can be expressed accordingly.
\begin{subequations}
\begin{align}
    \frac{\text{d}\mathcal{L}}{\text{d}{{a}_{i}}}=\sum\limits_{\text{n}=1}^{H}{\sum\limits_{m=1}^{W}{\frac{\text{d}\mathcal{L}}{\text{d}{{S}_{n,m}}}(-x{{_{i}^{n,m}}^{2}}\exp (-\delta^{n,m}_{j})}})
\end{align}
\begin{align}
    \frac{\text{d}\mathcal{L}}{\text{d}{{b}_{i}}}=\sum\limits_{\text{n}=1}^{H}{\sum\limits_{m=1}^{W}{\frac{\text{d}\mathcal{L}}{\text{d}{{S}_{n,m}}}(-x_{i}^{n,m}y_{i}^{n,m}\exp (-\delta^{n,m}_{j})}})
\end{align}
\begin{align}
     \frac{\text{d}\mathcal{L}}{\text{d}{{c}_{i}}}=\sum\limits_{\text{n}=1}^{H}{\sum\limits_{m=1}^{W}{\frac{\text{d}\mathcal{L}}{\text{d}{{S}_{n,m}}}(-y{{_{i}^{n,m}}^{2}}\exp (-\delta^{n,m}_{j})}})
\end{align}
\end{subequations}

\subsubsection{Gradient of Intensity Computation Process}
The calculation of scattering intensity is divided into two parts: the partial derivatives of the loss function $\mathcal{L}$ with respect to both the forward and backward scattering coefficients, and the partial derivatives of the loss function $\mathcal{L}$ with respect to the phase function $P$.
The derivative of the loss function $\mathcal{L}$ with respect to the phase function of the $i$-th Gaussian component is given by 
\begin{equation}
\begin{aligned}
\frac{\mathrm{d}\mathcal{L}}{\mathrm{d}P_i} = 
\sum_{p=1}^N \sum_{q=1}^M \sum_{a=i}^{G_{p,q}}
\frac{\mathrm{d}\mathcal{L}}{\mathrm{d}I_a} \frac{\mathrm{d}I_a}{\mathrm{d}P_i}.
\end{aligned}
\end{equation}
Since the partial derivative of the loss function 
$L$ with respect to the forward scattering coefficient$k_{e}^{-}$is the same as that with respect to the backward scattering coefficient$k_{e}^{+}$, we only present the derivative with respect to the backward scattering coefficient $k_{e}^{+}$ here, as shown below.
\begin{equation}
\begin{aligned}
\frac{\mathrm{d}\mathcal{L}}{\mathrm{d}k^+_{e_i}} = 
\sum_{p=1}^N \sum_{q=1}^M \sum_{a=i}^{G_{p,q}} 
\frac{\mathrm{d}\mathcal{L}}{\mathrm{d}I_a} 
\frac{\mathrm{d}I_a}{\mathrm{d}E_i^{+p,q}} 
\frac{\mathrm{d}E_i^{+p,q}}{\mathrm{d}k^+_{e_i}}.
\end{aligned}
\end{equation}

\subsubsection{Gradient of Gaussian Parameters}
The inverse of the covariance matrix contributing to the loss function can be decomposed into two parts: the two-dimensional covariance matrices in both the computational plane and the imaging plane each contribute to the derivative. Since the same orthographic projection is used in the projection process, only the projection Jacobian matrices differ. Thus, the gradient of the projection process is given by:
\begin{equation}
    \frac{\text{d}\mathcal{L}}{\text{d}\Sigma }=\frac{\text{dL}}{\text{d}{{\Sigma }^{\prime }}}\frac{\text{d}\Sigma _{{}}^{\prime }}{\text{d}{{\Sigma }_{{}}}}=\frac{\text{dL}}{\text{d}\Sigma _{{}}^{\prime}}\mathbf{R}\mathbf{J}{\mathbf{J}^{T}}{\mathbf{R}^{T}}
\end{equation}
The phase function is determined by the Gaussian point cloud center and the spherical harmonic coefficients. By computing the gradients of the loss function with respect to the Gaussian point cloud center $\frac{\text{d}\mathcal{L}}{\text{d}\mathbf{x}}$ and the spherical harmonic coefficients $\frac{\text{d}\mathcal{L}}{\text{dSH}}$, the point cloud position and the phase function can be optimized simultaneously.

\subsubsection{Densification and Pruning Operation}
To address regions with insufficient reconstruction accuracy, the densification operation adaptively duplicates Gaussian primitives based on gradient thresholds and displaces them along the positional gradient direction, thereby enhancing reconstruction in under-represented areas. For over-reconstructed regions, the operation splits oversized Gaussians and similarly displaces the subdivided primitives following the gradient direction. Furthermore, the process eliminates excessively large Gaussians or those with excessively small backscattering coefficients to reduce computational overhead and GPU memory consumption.

\section{Experimental Results}
The experiments in this study comprise three principal components: forward rendering simulation, target reconstruction experiments using both simulated and MSTAR datasets, and ablation studies. In the forward simulation experiments, we validated the effectiveness of the differentiable Gaussian rasterizer by generating SAR images of simplified architectural models, thereby confirming that the forward process adheres to radar imaging principles. The synthetic point cloud reconstruction experiments evaluated the SAR-GS method's capability in learning target geometry while verifying its model accuracy. For the MSTAR-based target reconstruction experiments, we systematically investigated the SAR-GS method's fitting capability under multi-view scattering characteristic discrepancies in real-world scenarios, along with its characteristic-preserving reconstruction performance. Finally, the ablation studies examined the impact of densification operations and initial point cloud quantity on target reconstruction, with quantitative evaluation of reconstruction performance under various conditions.

\subsection{Forward Rendering}
In the forward rendering experiments, canonical architectural models were geometrically simplified into rectangular prisms. The rationality of our SAR Differentiable Gaussian Rasterizer (SDGR) was validated through SAR image renderings across varying elevation and azimuth angles. Following SAR imaging's mapping and projection mechanisms, building manifestations in SAR imagery were systematically analyzed as composite phenomena comprising  scattering of ground (SG), scattering of wall(SW), scattering of roof (SR), and shadow (S) components.

The scattering energy distribution demonstrated explicit dependence on both radar elevation angles and architectural parameters - building height (h) and roof width (w). Under controlled experimental conditions (radar elevation angle: 45°).

\begin{figure}[htpb]
    \centering
    \includegraphics[width=1\linewidth]{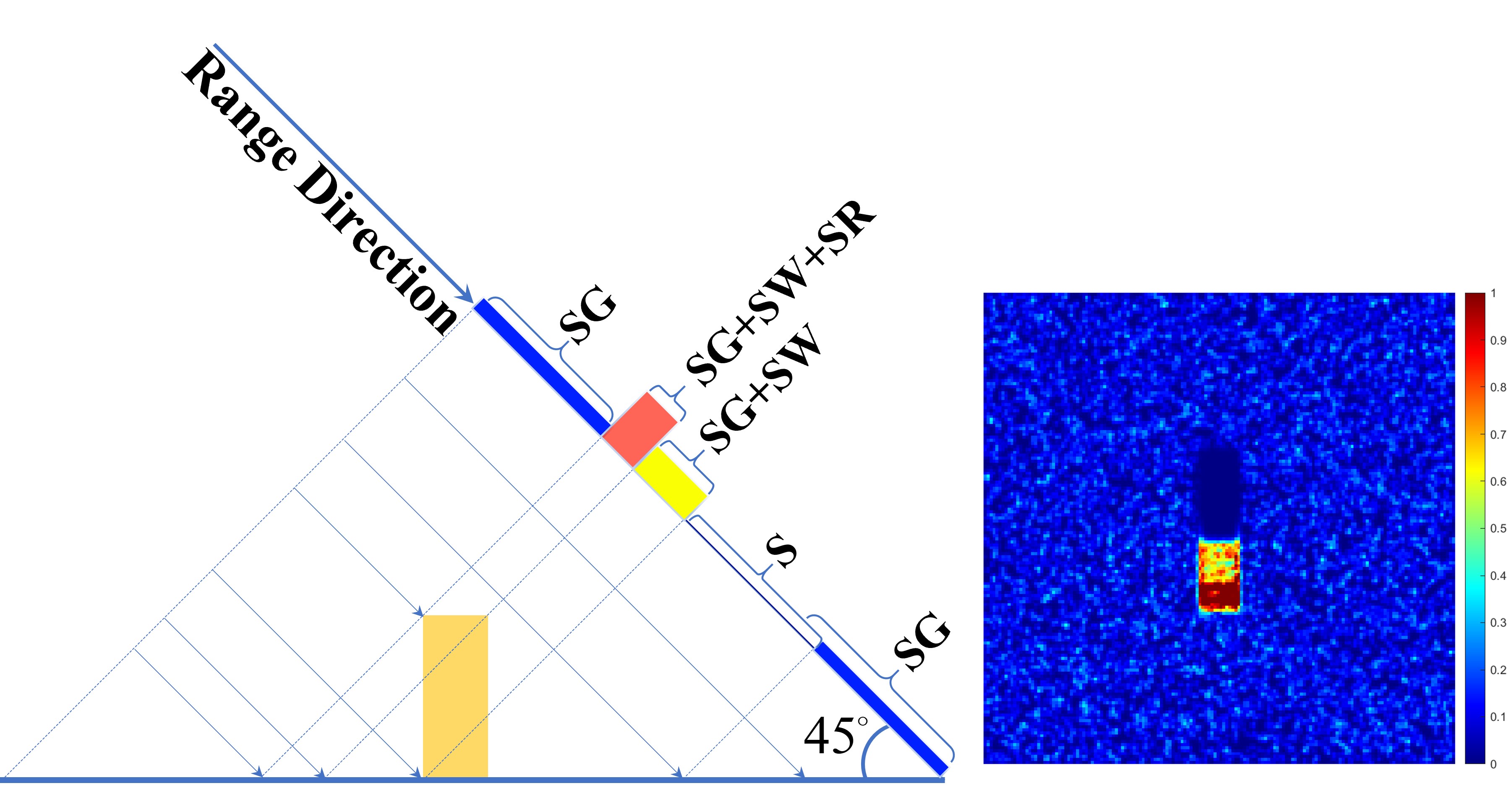}
    \caption{Architectural Model A: The roof width is smaller than the building's height.}
    \label{Architecture Model A}
\end{figure}

A building configuration with h$=$10m (building height) and w$=$4m (roof width) was established. When h $>$ 2w, the combined scattering distribution (SW+SR+SG) exceeded twice the magnitude of the SR+SG configuration, as depicted in Fig. \ref{Architecture Model A}. 

\begin{figure}[htpb]
    \centering
    \includegraphics[width=1\linewidth]{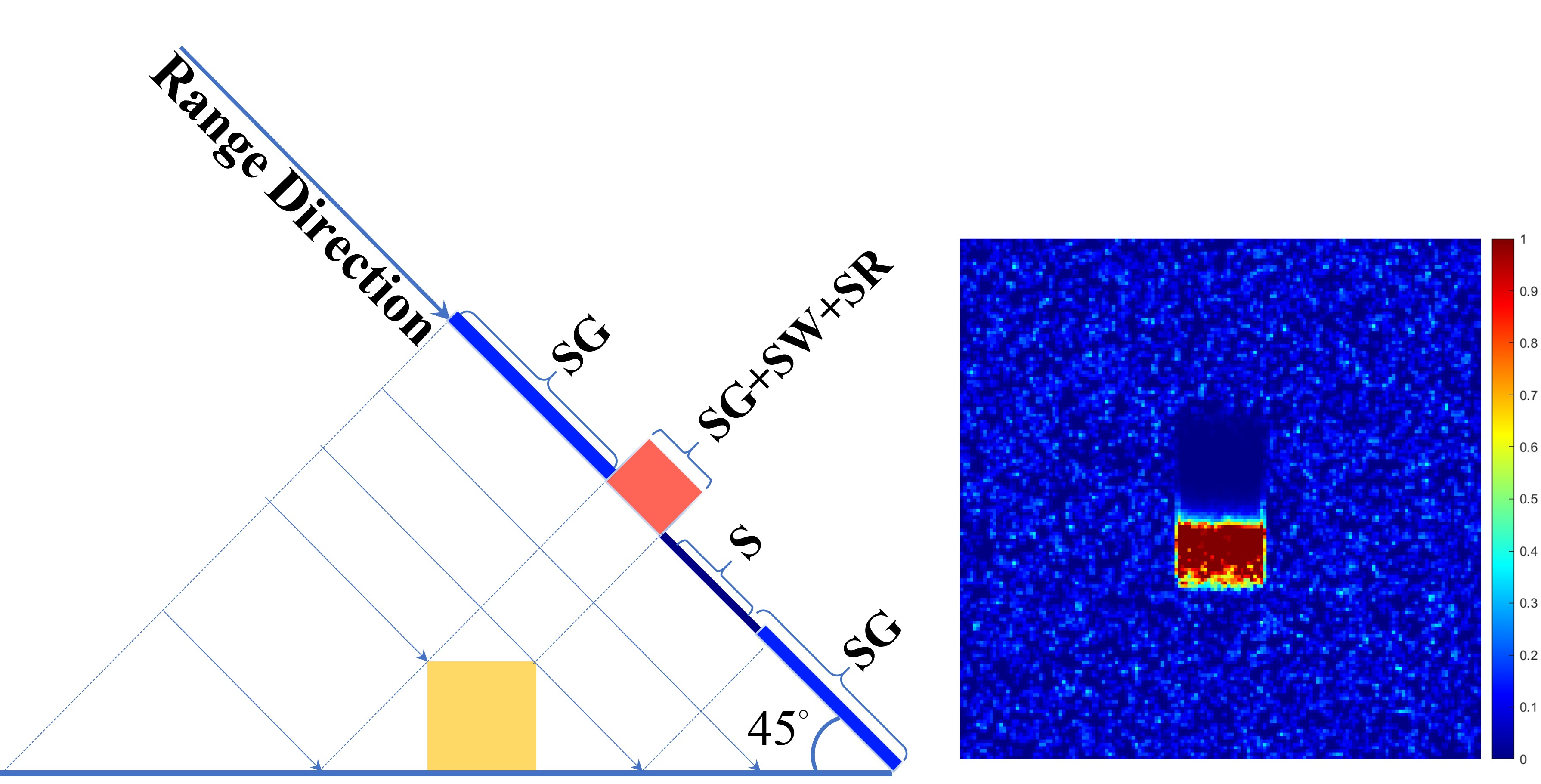}
    \caption{Architectural Model B: The roof width equals the building's height.}
    \label{Architecture Model B}
\end{figure}

For the case of h=8m and w=8m (i.e., h$=$w), the scattering regime exclusively exhibited SW+SR+SG components, with the shadow length precisely doubling the scattering distribution magnitude as depicted in Fig. \ref{Architecture Model B}). 

\begin{figure}[htpb]
    \centering
    \includegraphics[width=1\linewidth]{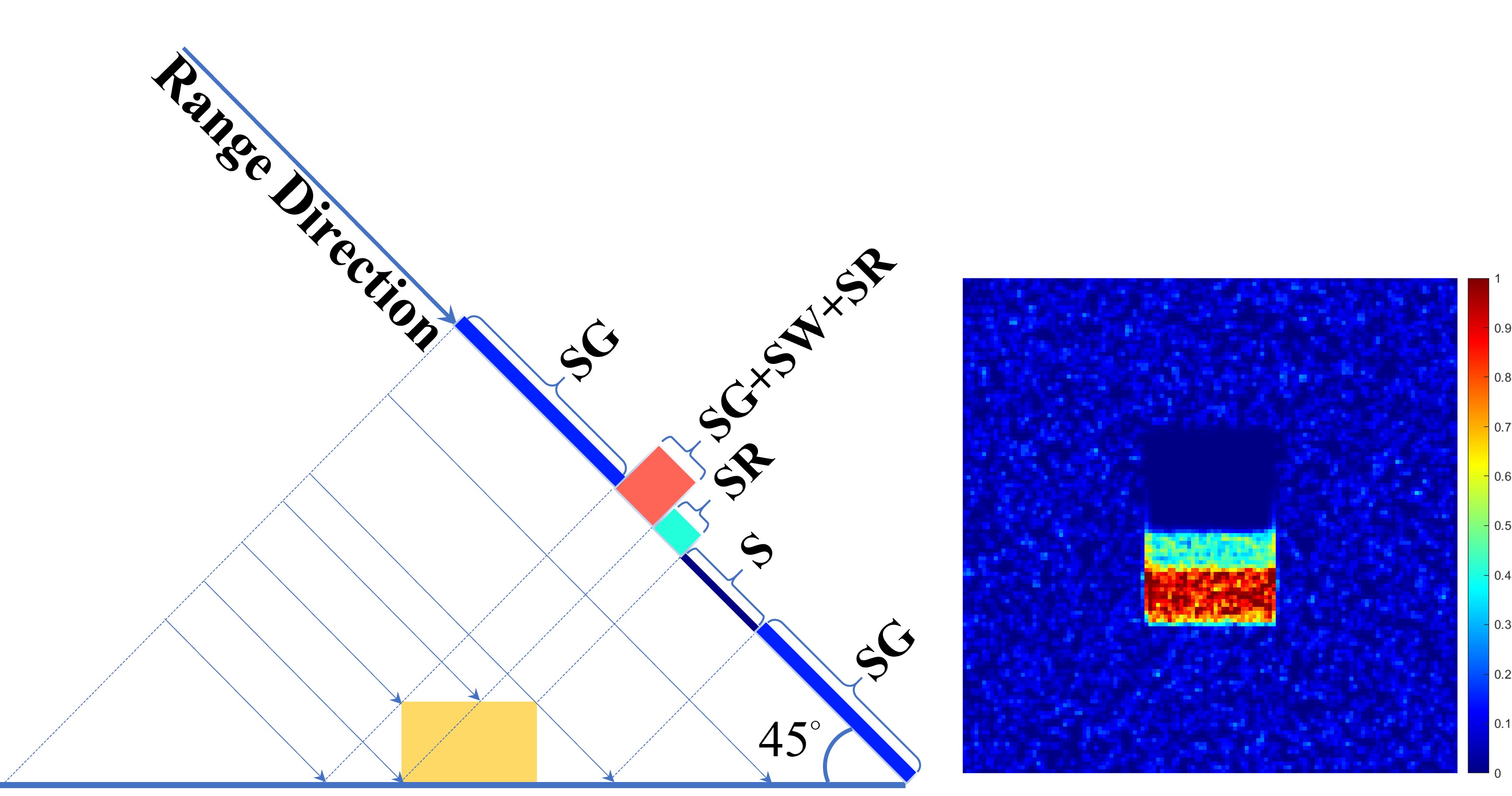}
    \caption{Architecture Model C: The roof width exceeds the building's height.}
    \label{Architecture Model C}
\end{figure}

Under the configuration h=6m and w=10m (i.e., h$<$w), the scattering signatures manifested a composite distribution comprising SW+SR+SG along with distinct SR components, as illustrated in Fig. \ref{Architecture Model C}.

\begin{figure}[htbp]
    \centering
    \subfigure[]{
    \includegraphics[width=0.45\linewidth]{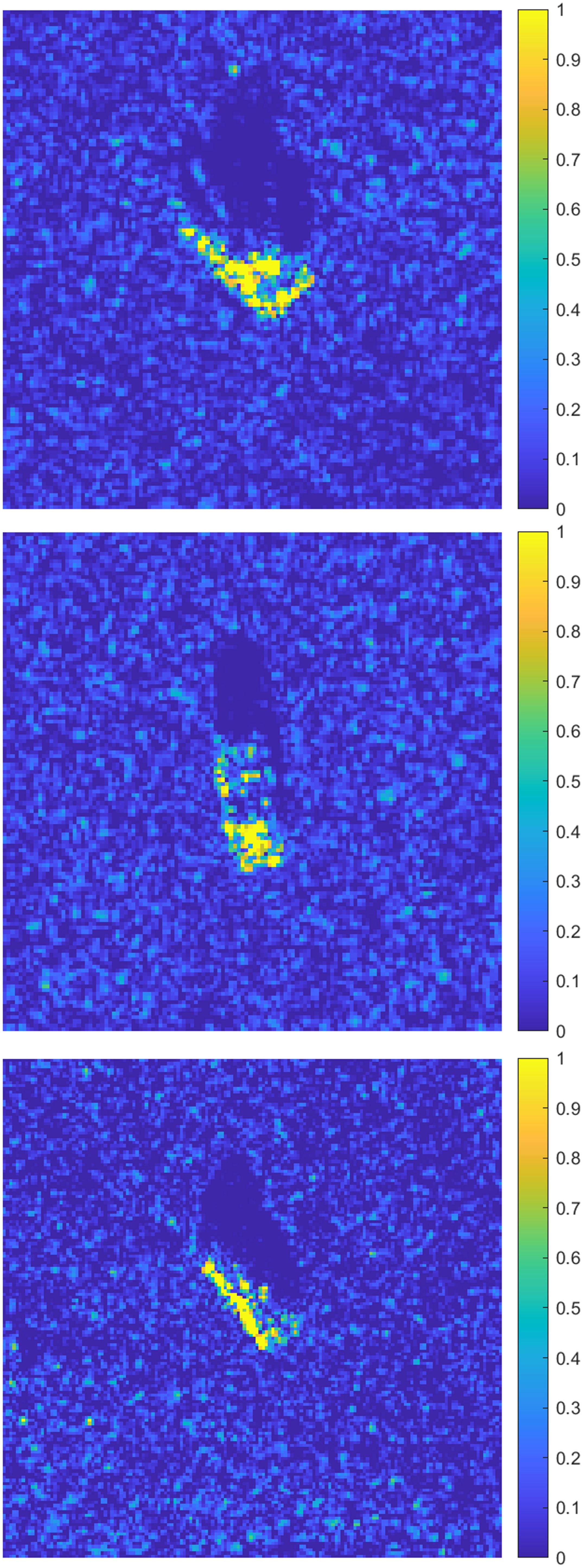}
    }
    \subfigure[]{
    \includegraphics[width=0.45\linewidth]{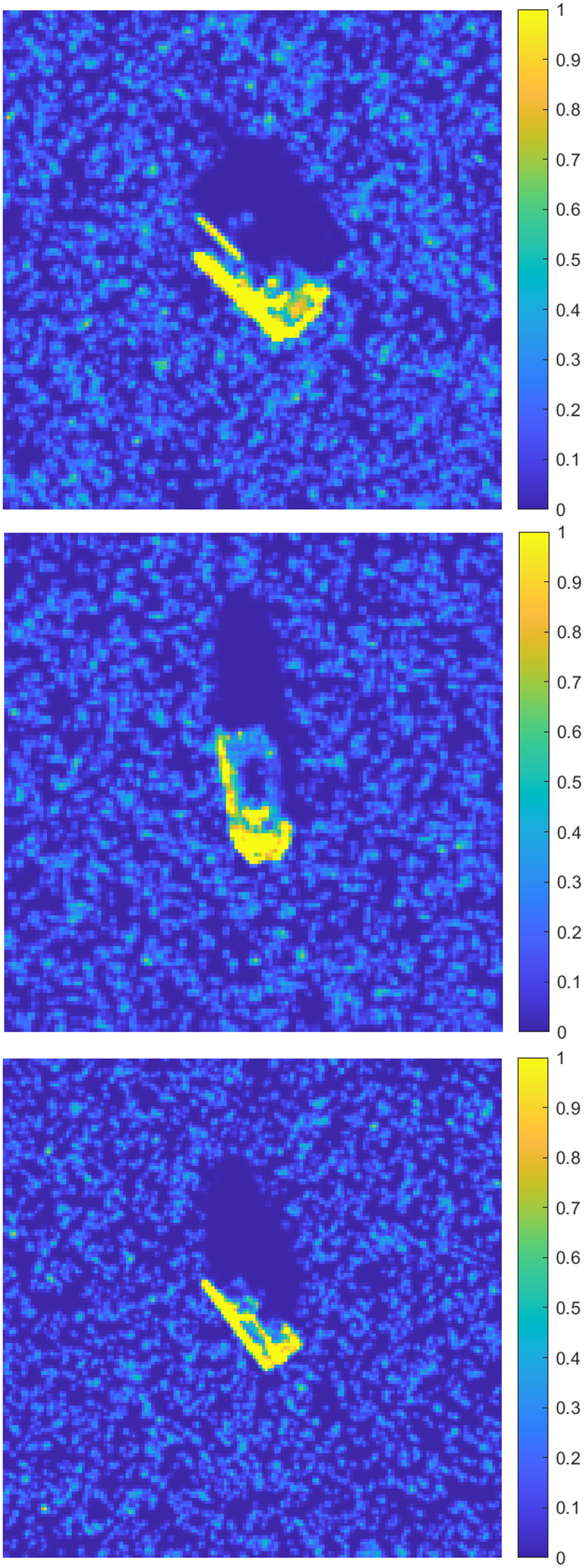}
    }\\
    \caption{Visual comparison between MSTAR(a) and rendered(b) SAR images(rendered by our method) of T72, BTR70, and 2S1 targets.}
    \label{fig:rendercontrast}
  \end{figure}

To further validate the rationality of SDGR's forward rendering process, we conducted rendering experiments on three kinds of vehicles (T72 , BTR70 and 2S1), with comparative analysis performed against corresponding SAR images from the MSTAR dataset,  as illustrated in Fig. \ref{fig:rendercontrast}. Comparative analysis with the MSTAR dataset demonstrates that our SDGR-simulated images exhibit high contour resemblance to the real SAR images, effectively replicating the shadow effects caused by radar beam occlusion from targets. Furthermore, the proposed method achieves satisfactory reconstruction fidelity for general rough surface backgrounds.


\subsection{Target Reconstruction based-on Rendered Dataset}
\subsubsection{Rendered Dataset}
In the point cloud reconstruction experiment, to preliminarily validate the reconstruction capability of SAR-GS for target point clouds. In this section of the experiment, we utilize forward-rendered SAR images derived from the SAR-GS method, including targets such as the T72, BTR80, and KRAZ. Prior to the reconstruction experiment, the SAR-GS method is employed to compute the backscattering intensity of Gaussian primitives in the scene, thereby generating multi-view SAR image sequences. During simulation, the incidence angles are set to 15°, 45°, and 75°, while the azimuth angles are sampled at 15° intervals. The simulation parameters are listed in the table. The resolution of the simulated data is 0.3 m in both the azimuth and range directions. To mitigate the influence of ground scattering on the target during reconstruction, sampling is performed exclusively on the target, excluding the ground. The resulting reference ground-truth point cloud is used as input for the forward renderer to generate the rendered dataset.


\begin{table}[!t]
    \caption{Parameters Setting of Rendered Dataset}\label{PARAMETERS SETTING OF SIMULATED DATASET}
    \centering
    \begin{tabular}{>{\rule[0ex]{0pt}{2ex}}c c c}
    \Xhline{1pt}  
    \multicolumn{2}{c}{Item} & Parameter  \\
    \hline
    \multicolumn{2}{c}{Target Category} & T72,BTR70,KRAZ\\
    \hline
    \multicolumn{2}{c}{Resolution}   & 0.3m  \\
    \hline
    \multirow{2}{*}{Azimuth Sampling}&Train   & $\{0^\circ,45^\circ,90^\circ,\ldots,315^\circ\}$  \\
    \cline{2-3}
    &Test  & $\{15^\circ,30^\circ,60^\circ,\ldots,345^\circ\}$\\
    \hline
    \multirow{2}{*}{Incident Sampling}&Train   & $\{15^\circ,45^\circ,75^\circ\}$  \\
    \cline{2-3}
    &Test  & $\{15^\circ,45^\circ,75^\circ\}$\\
    \Xhline{1pt}  
    \end{tabular}
\end{table}

\subsubsection{Evaluation Metrics}
To evaluate the performance of SAR-GS in target reconstruction tasks, we assess our approach through both image-based and point cloud reconstruction perspectives.

For the image-based evaluation, in addition to visual comparisons between multi-view generated images and ground truth images, we employ three quantitative metrics: Peak Signal-to-Noise Ratio (PSNR), Structural Similarity Index (SSIM), and Learned Perceptual Image Patch Similarity (LPIPS). The Peak Signal-to-Noise Ratio (PSNR) quantifies image quality by computing the Mean Squared Error (MSE) between the original and reconstructed images, followed by a logarithmic conversion of the ratio between the squared peak signal value (MAX) and MSE, expressed in decibels (dB). A higher PSNR value indicates lower distortion, as defined by:
\begin{equation}
    \mathrm{PSNR}=10\cdot\log_{10}\left(\frac{\mathrm{MAX}^2}{\mathrm{MSE}}\right)
\end{equation}
However, under SAR-specific evaluation conditions, PSNR may inadequately reflect the structural learning performance of targets in the presence of high-intensity background noise. To address this limitation, we incorporate SSIM and LPIPS as complementary metrics. SSIM evaluates image similarity by analyzing structural information, comparing luminance, contrast, and structural patterns between images to assess their overall similarity. 
\begin{equation}
    \mathrm{SSIM}(x,y)=\frac{(2\mu_x\mu_y+C_1)(2\sigma_{xy}+C_2)}{(\mu_x^2+\mu_y^2+C_1)(\sigma_x^2+\sigma_y^2+C_2)}
\end{equation}
LPIPS, in contrast, quantifies the perceptual distance between images using deep features, aligning more closely with human visual perception. In our study, we extract features via the AlexNet network to compute LPIPS values.
\begin{equation}
\begin{aligned}
    \mathrm{LPIPS}(x,y)=&\sum_lw_l\frac{\sum_{h,w,c}\left(F_l^{(x)}(h,w,c)  -F_l^{(y)}(h,w,c)\right)^2}{H_lW_lC_l} 
\end{aligned}
\end{equation}
This multi-metric framework ensures a comprehensive evaluation of reconstruction quality, balancing pixel-level fidelity (PSNR), structural coherence (SSIM), and perceptual consistency (LPIPS) under SAR-specific noise conditions.

For the evaluation of point cloud reconstruction performance, we adopt the Chamfer Distance (CD) and F1-Score as the quantitative metric. CD serves as a metric for quantifying the similarity between two point sets. It is computed as the average distance from each point in set A (e.g., the reference point cloud) to its nearest neighbor in set B (e.g., the reconstructed point cloud), plus the average distance from each point in set B to its nearest neighbor in set.
Formally, let A = R (reference/ground truth) and B = G (reconstructed point cloud). The Chamfer Distance is defined as:
\begin{equation}
\begin{aligned}
    \mathrm{CD}(A,B)=\frac{1}{2}\left(\sum_{a\in A}\min_{b\in B}\frac{\|a-b\|^2}{|A|}+\sum_{b\in B}\min_{a\in A}\frac{\|b-a\|^2}{|B|}\right)
\end{aligned}
\end{equation}

F1-Score is commonly employed as quantitative metrics to evaluate the similarity between a reconstructed point cloud G and a ground truth point cloud R. Their computation depends on matching strategies and error tolerance thresholds $\tau$.

\begin{subequations}
    \begin{align}
        \text{Precision} = \frac{|\text{TP}|}{|\text{TP}| + |\text{FP}|},
    \end{align}
    \begin{align}
        \text{Recall} = \frac{|\text{TP}|}{|\text{TP}| + |\text{FN}|},
    \end{align}
    \begin{align}
        F_1 = 2 \times \frac{\text{Precision} \times \text{Recall}}{\text{Precision} + \text{Recall}},
    \end{align}
\end{subequations}
where Precision quantifies the proportion of correctly matched points in the predicted point cloud relative to all predicted matches, while Recall measures the fraction of correctly matched points in the reference (ground truth) point cloud relative to all existing points in the reference.

\subsubsection{Experiment Set Up}
We subsequently conducted point cloud reconstruction tests on three simulated targets. For each target, under the observation condition of radar altitude at 10000 meters (default in this experiment) and three predefined pitch angles, we selected 8 training images from 24 azimuth images (sampled at 15° intervals from 0° to 360°) per pitch angle by selecting every third image (at 45° intervals), while the remaining 16 images served as the test set. This experiment was performed on an NVIDIA RTX 3060 Ti GPU with 8GB VRAM, and the training time for each target is documented in Tab. \ref{Model Rendering Metrics}.

The training process was conducted over 30,000 iterations with target-specific initialization of point clouds. Based on dimensional estimates, the initial point clouds contained 15,000 points for T72 and BTR70 targets, while the KRAZ vehicle target was initialized with 10,000 points. We adopted a hemispherical initialization scheme for the point clouds, where the model optimized their positions through gradient descent. The training pipeline incorporated a densification process similar to 3D-GS, involving splitting or pruning oversized Gaussian primitives during optimization. The densification operation commenced at the 10,000th iteration and was terminated at the 25,000th iteration.

Furthermore, we observed that spherical harmonics (SH) exhibit strong representational capacity for anisotropic properties. However, an excessively rapid increase in SH degree can adversely affect the learning of point cloud positions. To mitigate this, we extended the training duration for low-order spherical harmonics, incrementally increasing the SH degree by one every 2,500 iterations until reaching the third order. This configuration ensures that the point clouds first learn accurate initial positions while effectively focusing on ray attenuation characteristics.

\subsubsection{Experimental Evaluation}
First, we quantitatively evaluated the similarity between rendered images from reconstructed point clouds and ground-truth images at corresponding viewing angles under test set conditions. Fig. \ref{fig:rendercontrast} present a visual comparison of rendering images generated by SAR-GS, real test images, and images rendered by the SAR-NeRF method. SAR-NeRF demonstrates exceptional fitting capability in the image domain, effectively reconstructing fine details while maintaining multi-view consistency. We therefore selected SAR-NeRF as our baseline for image-domain comparison. Quantitative results in the accompanying table demonstrate that SAR-GS outperforms SAR-NeRF across multiple metrics including PSNR, LPIPS, and SSIM for target rendering. Our method accurately reproduces the scattering energy distribution in synthetic images while faithfully reconstructing target details. Notably, SAR-GS achieves superior performance in high-frequency detail reconstruction, demonstrating its advanced image reconstruction capabilities.

\begin{figure*}[htpb]
    \centering
    
    \includegraphics[width=1\linewidth]{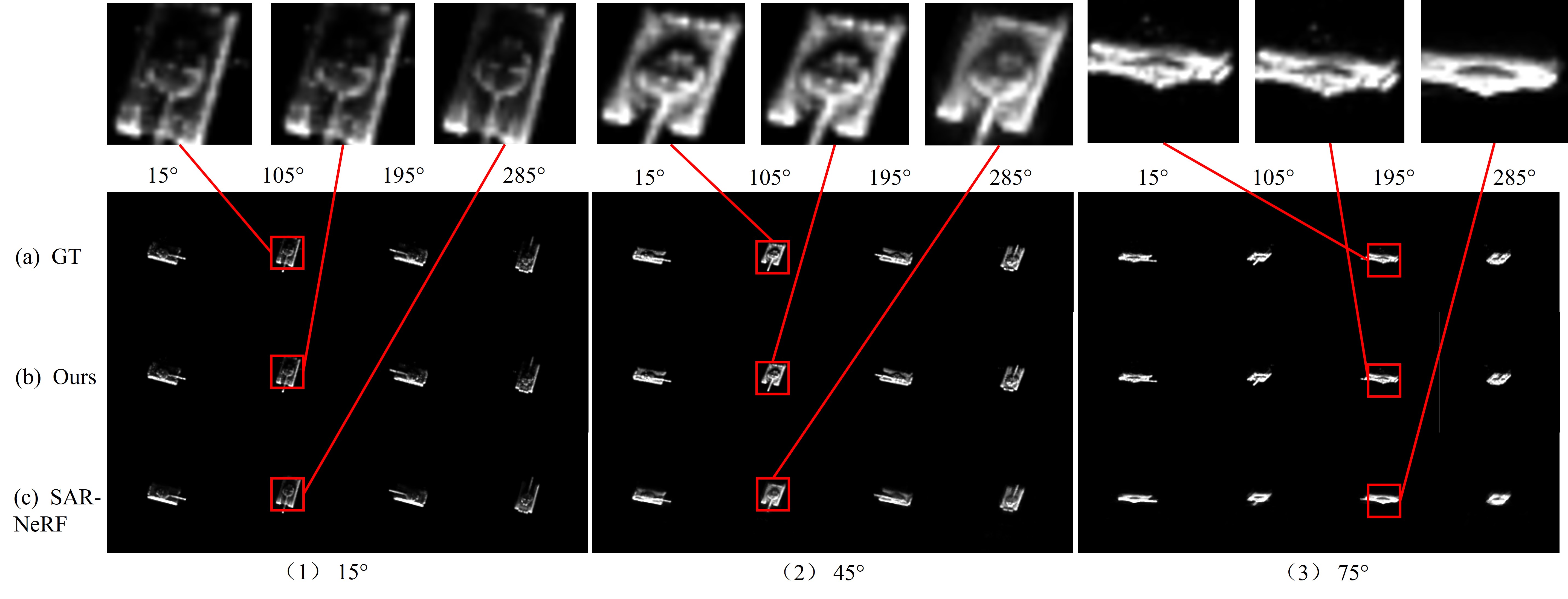}
    \caption{Comparative Visualization of T72 Target: Ground-Truth Rendered Images (a) vs. Our Method's Reconstructions (b) vs. SAR-NeRF Reconstructions (c). These include multi-azimuth rendered images at elevation angles of 15°, 45°, and 75°. From the image results, it can be observed thatthe proposed method exhibits superior reconstruction performance in terms of the high-frequency details of the target compared with SAR-NeRF. Specifically, SAR-NeRF yields smoother reconstruction results, while the images reconstructed by our method better restore the texture information.}
    \label{image_T72}

\end{figure*}

Subsequently, we evaluated the effectiveness of point cloud reconstruction using the SAR-GS method. Fig. \ref{image_T72} displays the reconstructed point clouds obtained from SAR image inputs when initialized with a hemispherical shell configuration. In the rendered dataset, our method demonstrates robust reconstruction performance. For instance, the reconstructed point clouds of T72 and BTR70 vehicle models clearly preserve geometric features of critical components such as gun barrels and turrets, while maintaining highly accurate overall contours that closely match the ground-truth point clouds. This precision confirms our method's capability to maintain key structural details while effectively suppressing noise artifacts inherent to SAR imaging. Furthermore, our 3D reconstruction metrics, including Chamfer Distance (CD) and F1-Score, achieve satisfactory results. These experimental findings validate the geometric fidelity of SAR-GS when reconstructing complex targets under sparse observation conditions.

\begin{figure*}
    \centering
    \includegraphics[width=1\linewidth]{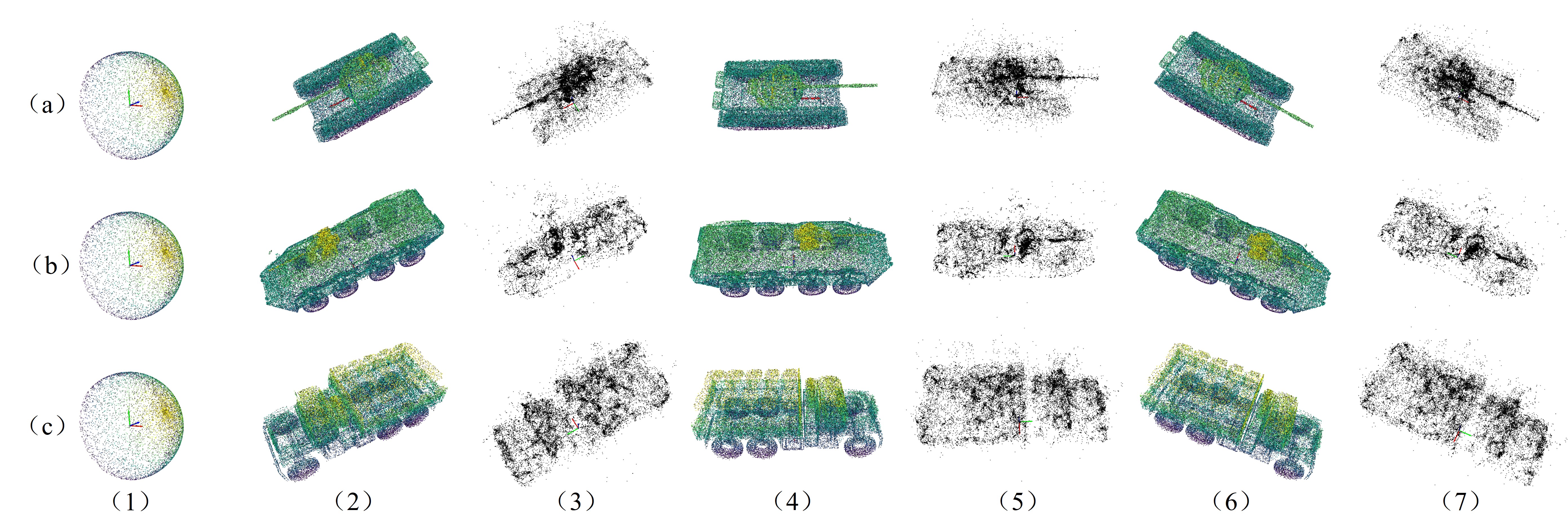}
    \caption{Reconstruction results for the three targets T72 (a), BTR80 (b) and KRAZ (c). Column (1) represents the initial point clouds of four targets. Columns (2), (4), and (6) display the ground-truth point clouds of the targets from three different angles. Correspondingly, columns (3),5), and (7) show the reconstructed point clouds at the same angles as the ground-truth point clouds presented in the preceding columns.}
    \label{Reconstruction_exSU-27}
\end{figure*}

\begin{table}[!t]
    \caption{Comparative Evaluation of Rendering Metrics Across Models}\label{Model Rendering Metrics}
    \centering
    \begin{tabular}{>{\rule[0ex]{0pt}{2ex}}c c c c c c}
    \Xhline{1pt}  
    Target & Method & SSIM$\uparrow$ & PSNR$\uparrow$ & LPIPS$\downarrow$  & Traing Time  \\
    \hline
    \multirow{2}{*}{T72}  & SAR-NeRF &   0.9894  & 34.29  & 0.0262  & 1h33min19s  \\
    \cline{2-6}
    & Ours & \textbf{0.9943}  & \textbf{35.91}  &\textbf{0.0165} & \textbf{6min40s}  \\
    \hline
    \multirow{2}{*}{BTR70}  & SAR-NeRF & 0.9911  & \textbf{34.54}  & 0.0224  & 1h34min26s  \\
    \cline{2-6}
    & Ours & \textbf{0.9928}  & 34.35  & \textbf{0.0176} & \textbf{5min58s}  \\
    \hline
    \multirow{2}{*}{KRAZ}  & SAR-NeRF & 0.9953 & 36.65  & 0.0178 & 1h30min32s  \\
    \cline{2-6}
    & Ours &  \textbf{0.9969} & \textbf{37.33}  & \textbf{0.0109}  & \textbf{6min16s}  \\
    \Xhline{1pt}  
    \end{tabular}
\end{table}


\begin{table}[!t]
    \caption{Performance Metrics of the Proposed Reconstruction Method}\label{Model Reconstruction Metrics}
    \centering
    \begin{tabular}{>{\rule[0ex]{0pt}{2ex}}c c c c c c c c}
    \Xhline{1pt}  
    Target & R$\rightarrow$G $\downarrow$ & G$\rightarrow$R $\downarrow$ & CD $\downarrow$ & P $\uparrow$  & R $\uparrow$ & F1 $\uparrow$  \\
    \hline
    T72  & 0.198 &   0.175  & 0.187  & 0.957  & 0.896& 0.925 \\
    \hline
    BTR70  & 0.236 & 0.220  & 0.228  & 0.876  & 0.819 &  0.846\\
    \hline
    KRAZ  & 0.206 & 0.208 & 0.207  & 0.934 & 0.800 & 0.862  \\
    \Xhline{1pt}  
    \end{tabular}
\end{table}


\subsection{Target Reconstruction based-on Real Dataset}

\subsubsection{Real Dataset}
To evaluate the effectiveness of the proposed method on real-world datasets, the MSTAR (Moving and Stationary Target Acquisition and Recognition)\cite{mstar_dataset} dataset was employed. As one of the most widely recognized public datasets in the field of synthetic aperture radar (SAR) imagery, the MSTAR dataset contains SAR images of multiple targets captured at various azimuth and depression angles. For the real-data experiments, T72 and BTR70 targets were selected as the reconstruction subjects. The reconstruction experiments were conducted using 8 SAR images with different azimuth angles, all acquired at a depression angle of 17°. The slant-range resolution and azimuth resolution of the images were both 0.3 meters.

\subsubsection{Evaluation Metrics}
To quantitatively validate the effectiveness of the proposed reconstruction algorithm, the same evaluation metrics as in the previous section were adopted: Chamfer Distance and F1-Score. These metrics were used to assess the quality of the reconstructed point clouds against the ground truth.

\subsubsection{Experiment Set Up}
We performed point cloud reconstruction experiments using real SAR data from the MSTAR dataset, focusing on T72, BTR70 and 2S1. The reconstruction was based on 8 SAR images with varying aspect angles (as illustrated in Fig) acquired at a depression angle of 73°, featuring consistent range and azimuth resolutions of 0.3m. 
\begin{figure}[htpb]
    \centering
    \includegraphics[width=1\linewidth]{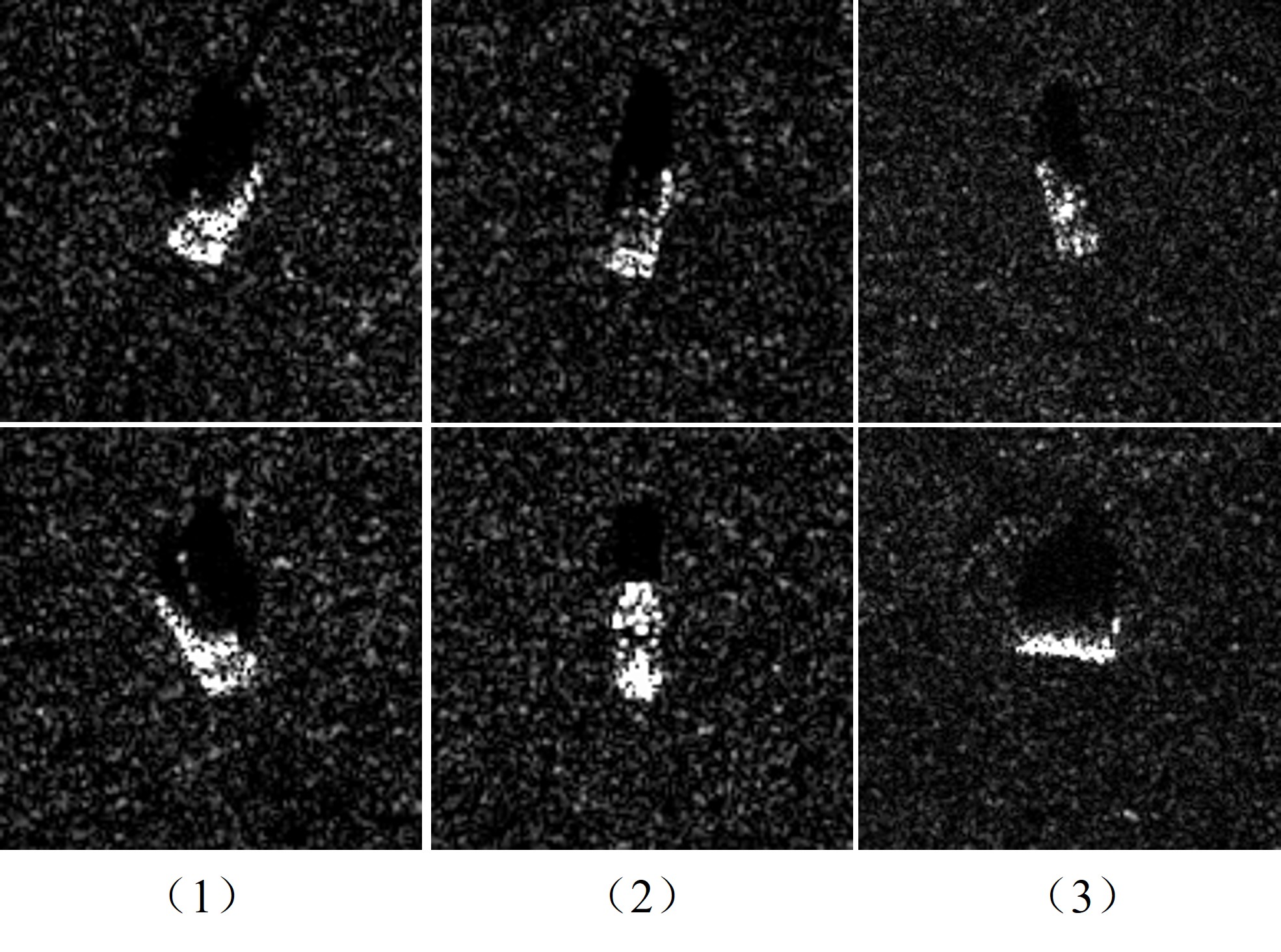}
    \caption{Training Dataset - SAR Images of T72(1), BTR70(2) and 2S1(3) }
    \label{train_set_T72.png}
\end{figure}

The training process employed 30,000 iterations with a hemispherical initialization of 60,000 points. Notably, we abandoned the densification approach commonly used with synthetic data due to its adverse effects on real SAR reconstruction. Our analysis revealed that gradient-based densification in real SAR scenarios leads to excessive Gaussian splitting in background regions, which ultimately obscures critical target scattering signatures. To address this, we implemented three key modifications: (1) strict bounding of Gaussian primitive displacements, (2) reduced learning rates for stable convergence, and (3) a controlled spherical harmonics progression scheme (increasing by one order every 2,500 iterations). This optimized configuration effectively preserves target scattering characteristics while maintaining geometric fidelity in the presence of complex background clutter.

\subsubsection{Experimental Results}
\begin{figure}[htbp]
    \centering
    
    \includegraphics[width=0.8\linewidth]{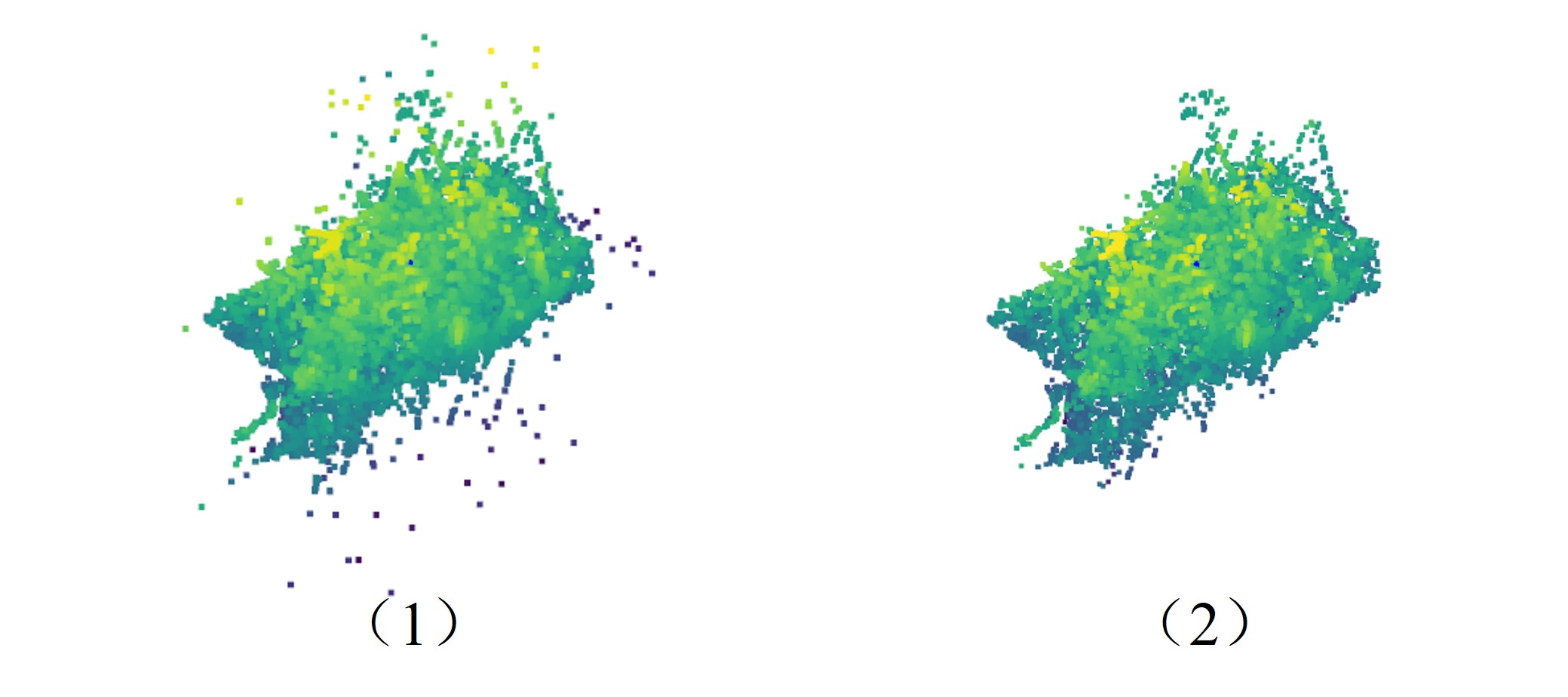}
    \caption{Comparison of point cloud results after clustering operation. (1) shows the reconstructed point cloud without clustering treatment, where floating point clouds are present around the target; (2) shows the point cloud after DBSCAN clustering.}
    \label{DBSCAN_filted}

\end{figure}

\begin{figure*}[htbp]
    \centering
    \includegraphics[width=1\linewidth]{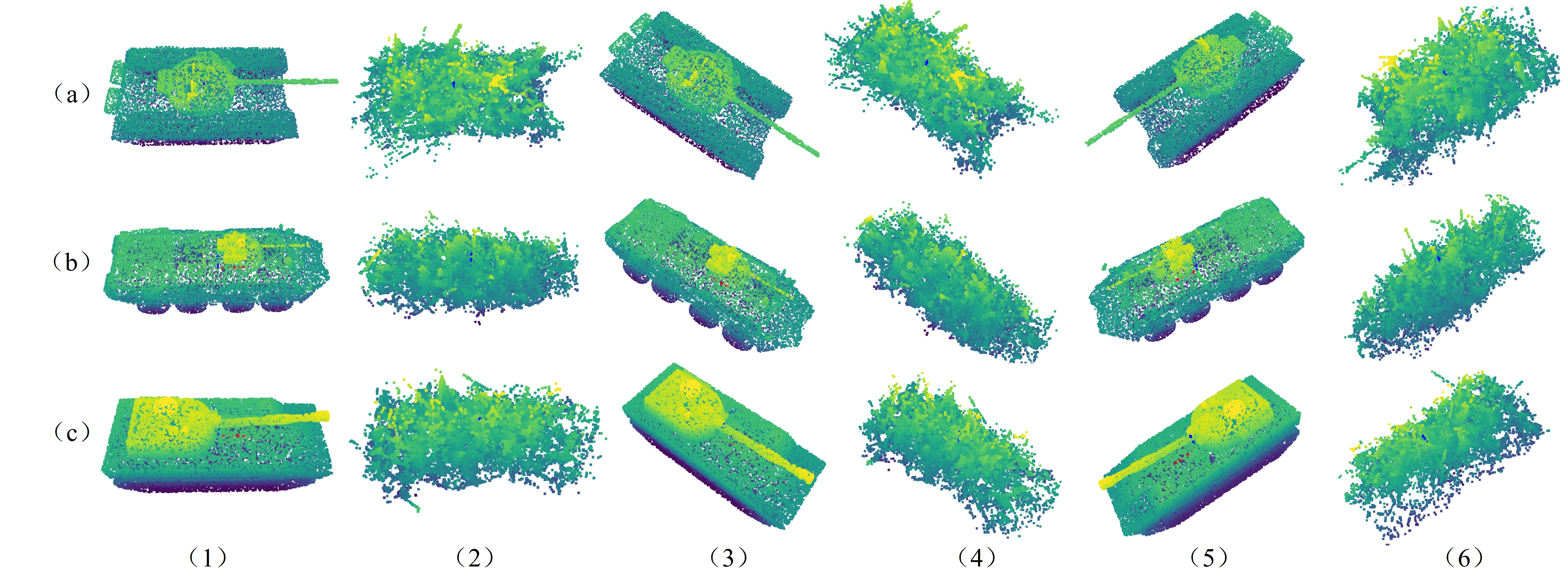}
    \caption{Comparison of Reference and Reconstructed Point Clouds for T72(a), BTR70(b) and 2S1(c) Vehicle Targets in MSTAR Measured Data. Column (1), (3), and (5) display the reference point clouds, while Column (2), (4), and (6) present the corresponding reconstructed point clouds.}
    \label{real-reconstruction}
\end{figure*}

The Fig. \ref{real-reconstruction} and tables present a comparison between the reconstructed point clouds generated by our method and the ground truth, along with the quantitative performance under the given experimental setup. The results demonstrate that the proposed method achieves satisfactory reconstruction performance, thereby preliminarily validating its capability for effective target reconstruction. The proposed method achieves an average training time of 16 minutes and 15 seconds for reconstructing a 60,000-point target under the experimental configuration, demonstrating a significant improvement in computational efficiency compared to SAR-NeRF, which typically requires several hours for model training. Although background noise adversely impacts reconstruction quality—specifically, as the model attempts to fit the noisy background, it consequently leads to the emergence of floating Gaussian components in the air.We employed the DBSCAN clustering method, which partitions regions with sufficient density into clusters by setting the maximum neighborhood distance and minimum number of neighborhood points. Points in low-density regions are identified as noise and filtered out, thereby eliminating non-target floating Gaussian points. Fig. \ref{DBSCAN_filted} presents the target reconstruction outcomes before and after clustering, where the point cloud was initialized with 70,000 points and no densification operation was applied.

\begin{table}[!t]
    \caption{Metrics of Reconstruction Performance on Real Datasets}\label{Metrics of Reconstruction Performance on Real Datasets}
    \centering
    \begin{tabular}{>{\rule[0ex]{0pt}{2ex}}c c c c c c c}
    \Xhline{1pt}  
    Target  &  R$\rightarrow$G $\downarrow$ & G$\rightarrow$R $\downarrow$ & CD $\downarrow$ & P $\uparrow$  & R $\uparrow$ & F $\uparrow$   \\
    \hline
    T72   & 0.121 & 0.186 & 0.154  & 0.986  & 0.910 & 0.946 \\
    \hline
    BTR70& 0.121 &   0.246  & 0.183 & 0.993  & 0.784 & 0.876 \\
    \hline
    2S1  & 0.129 & 0.189  & 0.159  & 0.993  & 0.904 &  0.946 \\
    \Xhline{1pt}  
    \end{tabular}
\end{table}

\begin{figure}[htpb]
    \centering
    \includegraphics[width=1\linewidth]{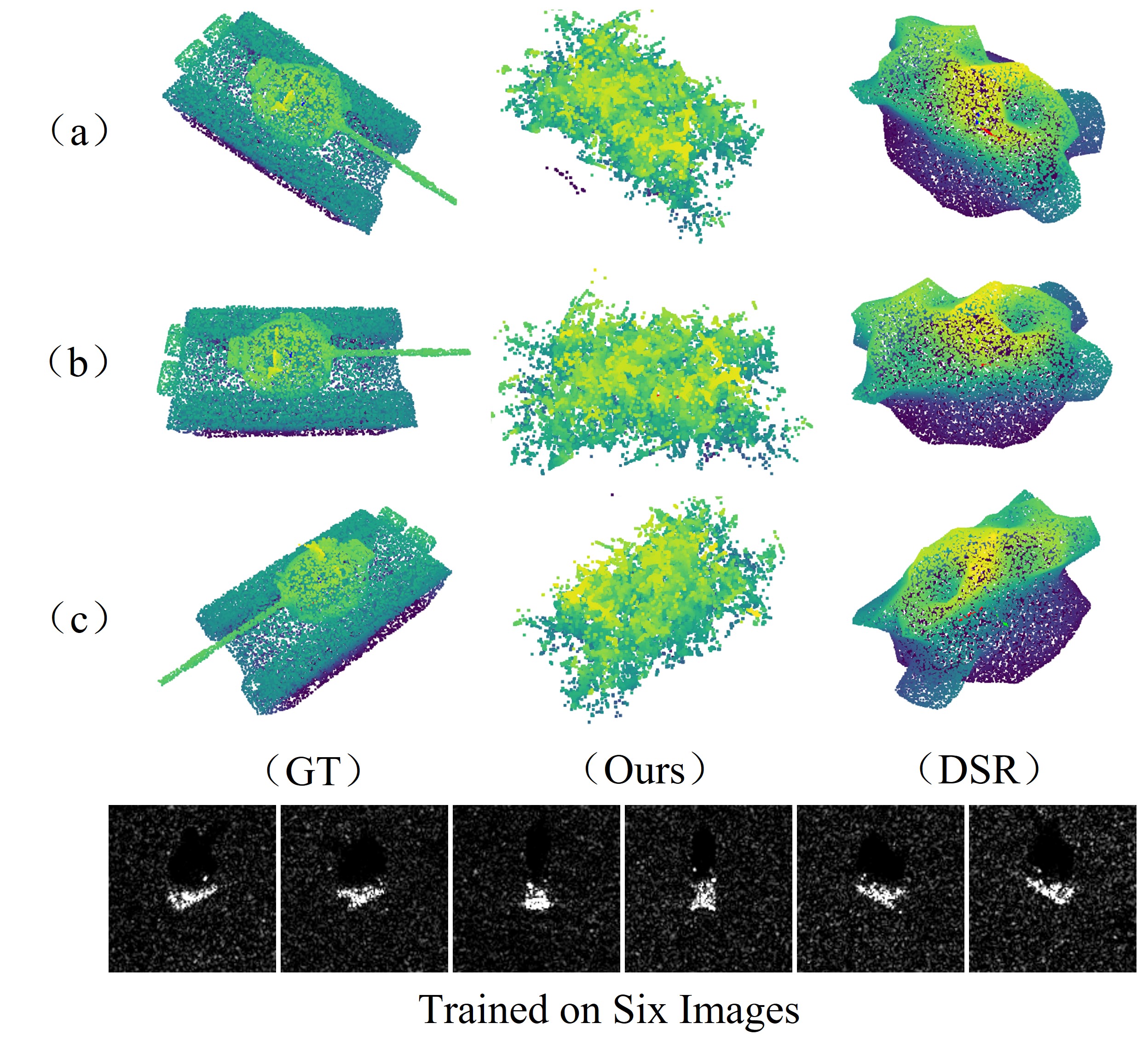}
    \caption{A Visual Comparison Between Our Method and the DSR Approach}
    \label{real_contrast_T72}
\end{figure}

We further conducted quantitative comparisons (as presented in Tab. \ref{COMPARISON RESULTS WITH DSR RECONSTRUCTION METHODS}) and visual comparisons (as illustrated in Fig. \ref{real_contrast_T72}) between the proposed method and the DSR algorithm \cite{9926979}. The specific operation is as follows: dense point clouds were uniformly sampled from the T72 mesh models generated by DSR, and evaluated using the same metrics (Chamfer Distance and F1-Score).
Experimental results demonstrate that, compared with DSR, the proposed method exhibits a slight decrease in coverage for target reconstruction but achieves higher precision. Notably, our method outperforms DSR in both metrics of average Chamfer Distance (CD) and F1-Score. Furthermore, under the same training equipment conditions (RTX 2080 Ti 11GB), the training time required for DSR is 9 minutes and 56 seconds, while the proposed method only takes 7 minutes and 20 seconds. Thus, the proposed method significantly reduces the training time, which highlights its superior practicality and effectiveness in real-world applications.


\begin{table}[!t]
    \caption{Comparison Results with DSR Reconstruction Method}\label{COMPARISON RESULTS WITH DSR RECONSTRUCTION METHODS}
    \centering
    \begin{tabular}{>{\rule[0ex]{0pt}{2ex}}c c c c c c c}
    \Xhline{1pt}  
    Method  & R$\rightarrow$G $\downarrow$ & G$\rightarrow$R $\downarrow$ & CD$\downarrow$& P $\uparrow$  & R $\uparrow$ & F$\uparrow$ \\
    \hline
    DSR  & 0.222 & \textbf{0.170} & 0.196 & 0.826  & \textbf{0.864} & 0.845  \\
    \hline
    Ours  &   \textbf{0.139}  & 0.217&  \textbf{0.179} & \textbf{0.980}  & 0.839 & \textbf{0.904} \\
    \Xhline{1pt}  
    \end{tabular}
\end{table}

\subsection{Ablation Study}

This section presents a comprehensive ablation analysis to evaluate two key parameters in our reconstruction pipeline: initial point cloud density and densification operation. The study systematically examines how these factors influence both quantitative metrics and qualitative reconstruction quality.

During the densification process, we adjusted the densification thresholds for SAR images: a gradient threshold of 0.003 was adopted, and the maximum radius of Gaussian primitives was restricted to 0.3 times the radar illumination range. Under this parameter configuration, the ablation experiment for the densification operation was completed.

\begin{figure}[htpb]
    \centering
    \includegraphics[width=1\linewidth]{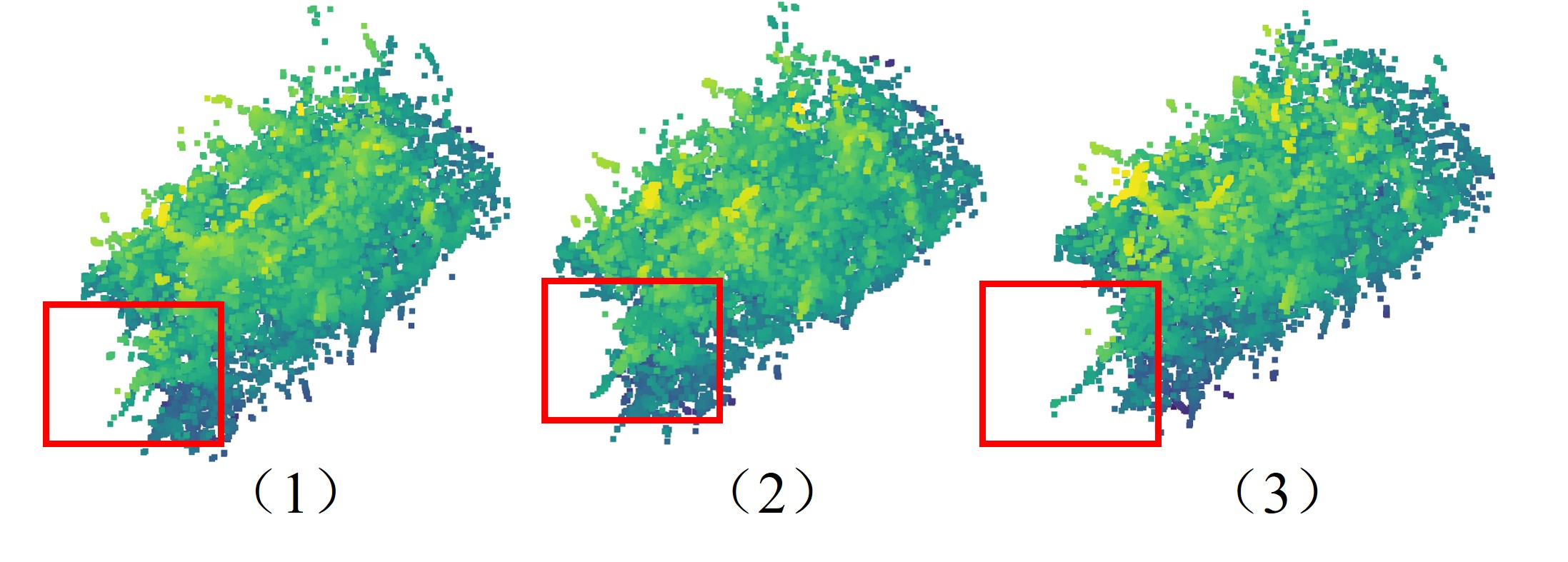}
    \caption{Reconstruction results under varying densification iterations: (1) Without densification, (2) 10,000 densification iterations, (3) 15,000 densification iterations.}
    \label{DenContrast}
\end{figure}

Regarding densification operations, we observe an interesting trade-off phenomenon. While excessive densification iterations negatively impact standard reconstruction metrics, our visual analysis reveals that moderate densification plays a crucial role in recovering fine geometric details. For instance, intermediate iterations successfully reconstruct subtle features like the characteristic inclined gun barrel of the T72 target\ref{DenContrast}, demonstrating that some level of densification is essential for accurate geometric representation.

\begin{table}[!t]
    \caption{Ablation Study of Densification Operation in Our Method}\label{Ablation Study of Densification Operation in Our Method}
    \centering
    \begin{tabular}{>{\rule[0ex]{0pt}{2ex}}c c c c c c c}
    \Xhline{1pt}  
    Iterations  & R$\rightarrow$G $\downarrow$ & G$\rightarrow$R $\downarrow$& CD$\downarrow$ & P$\uparrow$  & R $\uparrow$ & F$\uparrow$  \\
    \hline
    0 & \textbf{0.125} & \textbf{0.181} & \textbf{0.153} & 0.984 & \textbf{0.906} & 0.944\\
    \hline
    10000 & 0.129 & 0.184 & 0.157 & 0.987 & 0.905 & \textbf{0.944}\\
    \hline
    12000 & 0.127 & 0.190 &0.159 &\textbf{0.989}  & 0.892 &0.938\\
    \hline
    15000 & 0.132 & 0.187& 0.160 & 0.981  & 0.900 & 0.939 \\
    \hline
    18000 & 0.130 & 0.199& 0.165 & 0.987 & 0.890 &0.936 \\
    \hline
    20000 & 0.136 & 0.193& 0.164 & 0.984  & 0.895 & 0.938  \\
    \Xhline{1pt}  
    \end{tabular}
\end{table}

The investigation of initial point cloud density yields equally significant findings. Experimental results demonstrate that increasing the initial point count leads to measurable improvements in reconstruction accuracy, while incurring linearly scaled computational costs - as evidenced by the progressive extension of training time (9 minutes 20 seconds for 30,000 initial points versus 24 minutes 45 seconds for 110,000 initial points). This empirical evidence suggests that practitioners must carefully optimize the trade-off between reconstruction precision and computational resources in real-world deployments.


\begin{table}[!t]
    \caption{Ablation Study on Initial Point Cloud Density in Our Method}\label{Ablation Study on Initial Point Cloud Density in Our Method}
    \centering
    \begin{tabular}{>{\rule[0ex]{0pt}{2ex}}c c c c c c c}
    \Xhline{1pt}  
    Density  & R$\rightarrow$G $\downarrow$ & G$\rightarrow$R $\downarrow$& CD & P $\uparrow$  & R $\uparrow$ & F$\uparrow$  \\
    \hline
    30000 & 0.131 & 0.207& 0.169& \textbf{0.990} & 0.865 & 0.924 \\
    \hline
    50000 & 0.125 & \textbf{0.181}& 0.153&0.983  & 0.907& 0.944 \\
    \hline
    70000 & 0.119 & 0.189& 0.154& 0.986 & 0.907 & 0.945 \\
    \hline
    90000 & 0.117 & 0.184& 0.151 & 0.985 & \textbf{0.920}& 0.951 \\
    \hline
    110000 & \textbf{0.112} & 0.185 & \textbf{0.148}  & 0.987  & 0.918& \textbf{0.952}\\
    \Xhline{1pt}  
    \end{tabular}
\end{table}

These ablation results provide valuable practical guidance for implementation. They establish that while both parameters affect reconstruction quality, their impacts manifest differently - densification primarily influences geometric detail recovery while initial density affects overall reconstruction fidelity. The optimal configuration likely lies in balancing these factors based on application-specific needs and resource constraints.

\section{Conclusion} 
This paper proposes a novel SAR target reconstruction method based on Gaussian splatting, which integrates Mapping and Projection Algorithm with Gaussian splatting techniques to achieve effective learning of multi-aspect SAR image features and accurate point cloud reconstruction. Experimental results demonstrate that the proposed SAR-GS method exhibits several significant advantages over existing 3D reconstruction approaches. Compared with SAR-NeRF, our method achieves faster training speed and provides explicit target representation. Furthermore, unlike DSR methods that require manually annotated illumination and shadow maps as supervision, our approach significantly reduces data processing costs. More importantly, Gaussian point clouds, as a novel primitive for SAR scene representation, not only establish their effectiveness in SAR 3D reconstruction but also provide a solid foundation for various downstream tasks.

Nevertheless, the SAR-GS method still faces several limitations and challenges. First, it demands high-quality training samples, particularly requiring multi-depression angle data to ensure reconstruction accuracy. Second, the method remains susceptible to noise interference in SAR imagery, as evidenced by reconstruction difficulties caused by high-frequency background clutter in real-data experiments. Additionally, the lack of inter-point correlation during optimization leads to discrete point distributions and poor structural coherence in reconstructed point clouds. Future research will focus on establishing correlations between point clouds to generate more structured point cloud organizations, while simultaneously mitigating the impact of SAR image noise on reconstruction quality. These improvements will constitute key directions for advancing this field.

\bibliographystyle{IEEEtran}
\bibliography{reference}
\begin{IEEEbiography}[{\includegraphics[width=1in,height=1.25in,clip,keepaspectratio]{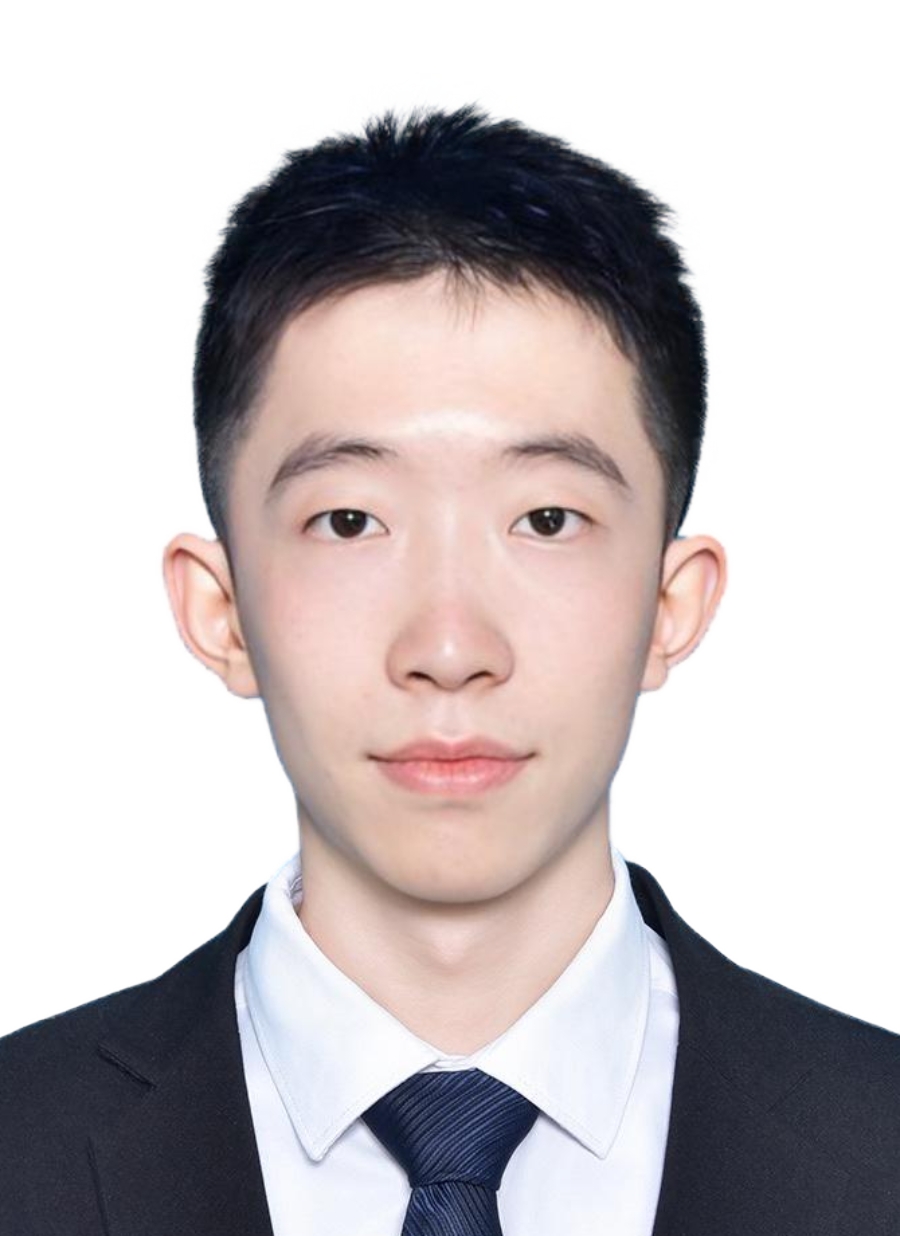}}]{Aobo Li} (Graduate Student Member, IEEE) received his Bachelor of Engineering (B.E.) degree from the School of Information Science and Engineering, Fudan University, in 2023. Currently, he is pursuing his Master of Engineering (M.E.) degree in the field of Electronic Information at the Key Laboratory of Electromagnetic Wave Information Science, Fudan University, Shanghai, China. His research interests include scattering modeling, Gaussian splatting, and 3D reconstruction algorithms.
\end{IEEEbiography}

\begin{IEEEbiography}[{\includegraphics[width=1in,height=1.25in,clip,keepaspectratio]{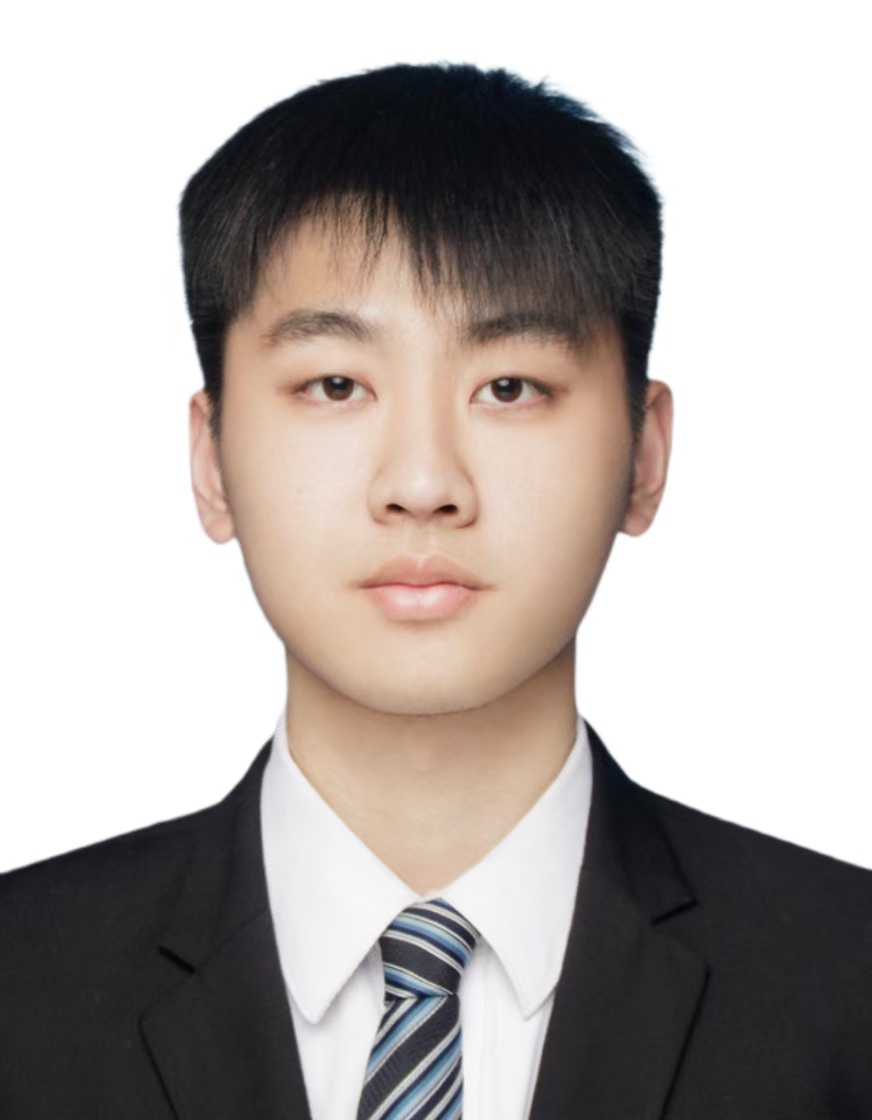}}]{Zhengxin Lei} (Graduate Student Member, IEEE) received the B.E. degree in communication engineering from Hubei University of Technology in 2019 and the M.E. degree in electronics and information engineering from Fudan University in 2022. He is currently pursuing his Ph.D. degree in electronic information at the Key Laboratory of Electromagnetic Wave Information Science, Fudan University, Shanghai, China. 
    His research interests include scattering modeling, Neural Radiance Fields.
\end{IEEEbiography}

\begin{IEEEbiography}[{\includegraphics[width=1in,height=1.25in,clip,keepaspectratio]{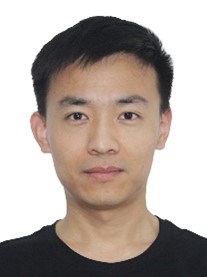}}]{Jiangtao Wei} (Member, IEEE) received the B.E. degree in weapons science and technology from North University of China, Taiyuan, China, in 2017 and the M.S. degree in aircraft design from Northwestern Polytechnical University, Xi'an, China, in 2021, and the Ph.D. in electronic information from the Key Laboratory of Electromagnetic Wave Information Science at Fudan University, Shanghai, China, in 2025. Since 2025, he has been an assistant researcher at the School of Future Information Innovation at Fudan University. His research interests include differentiable scattering modeling, SAR  inverse problems, and 3D reconstruction.
\end{IEEEbiography}

\begin{IEEEbiography}[{\includegraphics[width=1in,height=1.25in,clip,keepaspectratio]{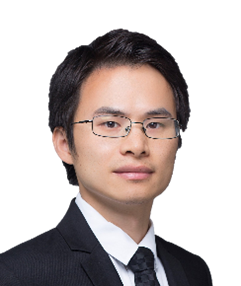}}]{Feng Xu} (Senior Member, IEEE) received the B.E. degree (Hons.) in information engineering from Southeast University, Nanjing, China, in 2003, and the Ph.D. degree (Hons.) in electronic engineering from Fudan University, Shanghai, China, in 2008. From 2008 to 2010, he was a Post-Doctoral Fellow with the NOAA Center for Satellite Application and Research (STAR), Camp Springs, MD, USA. From 2010 to 2013, he was a Research Scientist with Intelligent Automation Inc., Rockville, MD, USA. Since 2013, he has been a Professor with the School of Information Science and Technology, Fudan University. He is currently the Vice-Dean of the School of Information Science and Technology and the Director of the MoE Key Laboratory for Information Science of Electromagnetic Waves, Fudan University. He has published more than 60 articles in peer-reviewed journals or coauthored three books, among many conference papers and patents. His research interests include electromagnetic scattering theory, synthetic aperture radar (SAR) information retrieval, and advanced radar systems. Dr. Xu was a recipient of the Early Career Award of the IEEE Geoscience and Remote Sensing Society in 2014 and the SUMMA Graduate Fellowship in the advanced electromagnetics area in 2007. Among other honors, he was awarded the Second-Class National Nature Science Award of China in 2011. He is also the Founding Chair of the IEEE GRSS Shanghai Chapter and an GRSS AdCom Member. He also served as an Associate Editor for IEEE GEOSCIENCE AND REMOTE SENSING LETTERS.Neural Radiance Fields.
\end{IEEEbiography}

\end{document}